\documentclass[a4paper,11pt]{article}
\usepackage{jinstpub} 

\usepackage{amssymb}
\usepackage{graphicx}
\usepackage{caption}
\usepackage{subcaption}
\usepackage{siunitx}
\usepackage{lineno}
\usepackage{xcolor}
\newcommand{\added}[1]{#1}                    

\title{\boldmath Signal Processing and Machine Learning Algorithms for Precise Timing with PICOSEC Micromegas Detectors}

\author[a,b,1,2]{A. Kallitsopoulou\note{Corresponding author.}\note{Now at CEA-IRFU Université Paris-Saclay, Gif-sur-Yvette, France.}}
\author[a,b,3]{I. Maniatis\note{Now at Department of Particle Physics and Astronomy, Weizmann Institute of Science, Rehovot, Israel.}}
\author[c]{I. Manthos}
\author[d,e]{T. Papaevangelou}
\author[d,e,4]{L. Sohl\note{Now at TUV NORD EnSys GmbH \& Co. KG.}}
\author[a,b]{A. Tsiamis}
\author[a,b]{S.E. Tzamarias}

\affiliation[a]{Department of Physics, Aristotle University of Thessaloniki, University Campus, GR-54124, Thessaloniki, Greece}
\affiliation[b]{Centre for Interdisciplinary Research and Innovation (CIRI-AUTH), GR-57001, Thessaloniki, Greece}
\affiliation[c]{Institute of Experimental Physics, University of Hamburg, Luruper Chaussee 149, 22761, Hamburg, Germany}
\affiliation[d]{Université Paris-Saclay, F-91191, Gif-sur-Yvette, France}
\affiliation[e]{Commissariat à l'énergie atomique et aux énergies alternatives (CEA) - IRFU, F-91191, Gif-sur-Yvette, France}

\emailAdd{alexandra.kallitsopoulou@cea.fr}

\abstract{High particle rates in current and future experiments make pile-up phenomena a critical issue for extracting useful information. In this context, timing serves as a crucial fourth dimension for triggering and event reconstruction. The PICOSEC Micromegas detector has demonstrated precise timing capabilities on the order of tens of picoseconds. In this work, we develop and evaluate novel signal-processing algorithms to demonstrate the detector's potential for online precise timing. In particular, we propose an algorithm based on Artificial Neural Networks (ANNs), trained using a physically motivated model. The performance of the different algorithms is assessed using experimental data recorded during laser beam tests at the IRAMIS Facility of CEA Saclay. \added{A timing resolution of $18.3 \pm 0.6~\mathrm{ps}$, corresponding to $7.8$ photoelectrons in the gas volume}, is achieved---comparable to results obtained with the standard Constant Fraction Discrimination (CFD) technique. Additionally, we present an alternative algorithm that uses the integrated charge of the pulse exceeding a defined threshold as a parameter to correct for systematic effects.}


\begin{document}
\maketitle
\flushbottom

\section{Introduction}\label{sec:introduction}
\begin{figure}[hbt!]
	\begin{subfigure}{0.5\textwidth}
		\centering 
		\includegraphics[width=1.0\textwidth]{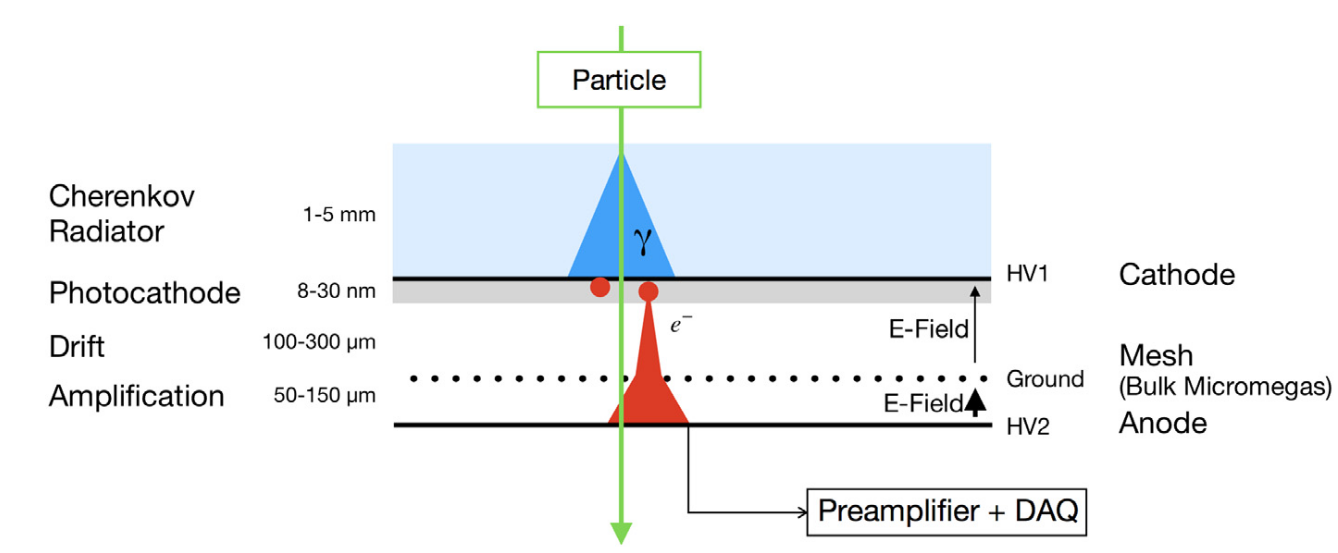}
		\caption{}
		\label{fig:PICOSECwaveform}
	\end{subfigure}
	\begin{subfigure}{0.5\textwidth}
		\centering 
		\includegraphics[width=0.7\textwidth]{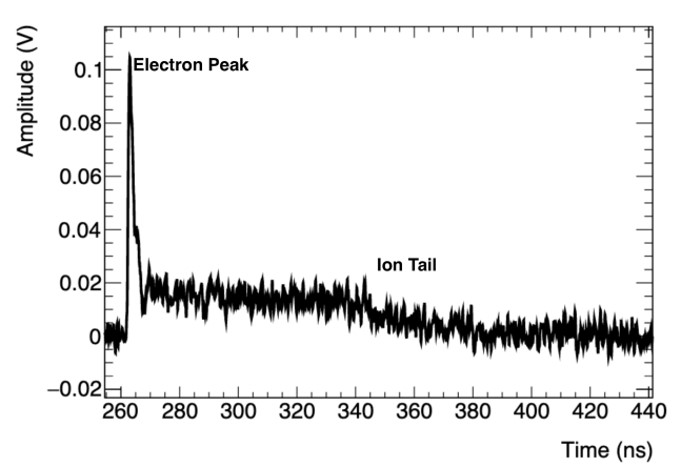}
		\caption{}
		\label{fig:picosec}
	\end{subfigure}
	\caption{(a) A graphical representation of a PICOSEC MicroMegas detector \cite{BORTFELDT2018317}. (b) The typical pulse of a PICOSEC-MM Detector is a two-component signal. It comprises a fast e-peak and an extended ion tail of the slow-moving ions.}
\end{figure}

The PICOSEC detector, shown in Fig.\ref{fig:PICOSECwaveform}, is based on the Micromegas technology \cite{GIOMATARIS}(hereafter called PICOSEC-MM), and includes a Cherenkov radiator and a photocathode. This design allows to an incoming relativistic charged particle to create Cherenkov photons inside the radiator, which convert to prompt photoelectrons in the photocathode. Additionally, the drift region has been reduced to the order of $\SI{200}{\micro\meter}$, compared to a typical value of 3-5\,mm to decrease the probability of direct ionization in the active volume, whilst also allowing for pre-amplification avalanche in the drift region due to the increased electric field.

\added{The data used for this study were collected at the Saclay Laser–Matter Interaction Center (IRAMIS/SLIC, CEA) using the LYDIL laser system~\cite{BORTFELDT2021165049,BORTFELDT2018317}. The facility employs a mode-locked Ti:Sapphire laser (MIRA 900 \footnote{Coherent. Mira Optima 900 F Data Sheet (2002)}, pumped by VERDI V10\footnote{Coherent. Verdi Family Data Sheet (2011)}) delivering pulses of $\sim120~\mathrm{fs}$ (FWHM) duration at 800\,nm with a 76\,MHz repetition rate. The wavelength was subsequently converted to 265\,nm through nonlinear frequency conversion using a barium borate (BBO) crystal, yielding ultraviolet pulses of approximately 150–200\,fs (FWHM) at the photocathode.}
The measurements were performed using an optimized PICOSEC-MM prototype with $\SI{119}{\micro\meter}$ drift gap,  operated at a high drift field and filled with $Ne:CF_4:C_2H_6$ (80:10:10) gas mixture. The electric field for these tests was set to 44\,kV/cm for the drift and 21\,kV/cm for the amplification detector's volumes.  The photocathode consisted of a thin film (18\,nm) of CsI deposited on a 10\,nm Aluminum layer, serving as the cathode. The laser beam was split, with one part directed at the PICOSEC-MM detector and the other illuminating a high-precision photodiode, with  \(\sim \) 3\,ps timing resolution, serving as a timing reference. 
The light beam that reached the detector was properly attenuated to control the number of photons hitting the PICOSEC photocathode, allowing us to study the detector's response to either a single-photoelectron or few, prompt photoelectrons \cite{sohl:tel-03167728}. The Signal Arrival Time (SAT) of the PICOSEC-MM was measured relative to the photodiode's signal on an event-by-event basis (hereafter called reference signal). Previous studies have shown that this optimized PICOSEC-MM detector has the capability of timing the arrival of a single photon with an accuracy better than 50\,ps \cite{Sohl_2020}.

In this report, novel timing methods are being developed and evaluated using data sets collected with the optimized PICOSEC-MM detector. The arrival time of a PICOSEC-MM signal (Fig.\ref{fig:picosec}) is typically extracted by analyzing the fully digitized electron peak waveform (e-peak), i.e. by fitting a logistic function to the leading edge of the e-peak. The SAT was then defined at the 20$\%$ of the e-peak maximum relative to the time-reference device signal (hereafter, this procedure is called full offline analysis, and the arrival time of the time-reference device is called reference time). Finally, the detector's timing resolution was defined as the RMS of the SAT distribution.


Two types of data sets were used: the SPE-set, comprising waveforms of single photoelectrons ( approximately 16k waveforms dataset), and the EXP-set (approximately 12k waveforms dataset), composed of waveforms of multiple photoelectrons. All waveforms were digitized by a fast (20\,GSamples/s), high-bandwidth oscilloscope (2.5\,GHz) after passing through a CIVIDEC current-sensitive preamplifier. Section \ref{sec:qup} first introduces an alternative timing method, based on the time-over-threshold technique. This algorithm uses the integral of the pulse above multiple thresholds (i.e. the equivalent of the charge) to correct for systematics. Building on the insights gained from these analytical signal processing approaches, the SPE-set is then used to emulate PICOSEC-MM signals with multiple photoelectrons, as described in Section \ref{sec:simulation}. After evaluating the performance of this model in Section \ref{sec:evaluation}, it is used to train an Artificial Neural Network (ANN) for timing the arrival of PICOSEC-MM waveforms, as detailed in Section \ref{sec:training}. In all methods discussed, the performance of the timing algorithms is assessed using the EXP-set waveforms.

It should be noted that the ANN-based timing technique performs optimally when the PICOSEC signal is relatively unaffected by time-walk. Due to the thin gap design, the prototype used in this study has a minimal time-walk effect, typically of only a few picoseconds \cite{utrobicic2024singlechannelpicosecmicromegas}. However, in cases where the signals experience more significant time-walk dependencies, the technique is no longer a good approximation, potentially degrading the timing resolution.

\section{Alternative Timing Technique with Constant Threshold}\label{sec:qup}
In our effort to propose an efficient timing technique, for scalable electronic devices with the PICOSEC-MM and reduce the cost of the experimental set-up, we evaluated the Constant Fraction method \cite{georgakopoulou2018100} on extracting the timing resolution of the PICOSEC-MM detector, by using the minimum information required. The goal of the developed signal processing algorithm is to retain the timing results achieved by the Constant Threshold method that takes into account the full waveform. In this approach, instead of using the time over threshold measurements, we use the integrals of the e-peak when exceeding certain thresholds. Hereafter such integrals will be referred to as \emph{Charge over Threshold} and symbolized as \emph{$Q_{up}$}.  This technique, like the Time over Threshold, suffers from systematics, introduced by the strong dependence of SAT on the pulse height. We prove that we can extract the experimental information that can to correct this effect, achieving precise timing accuracy \cite{kallitsopoulou2021development, Manthos_2021}. 

This technique requires thresholds adjusted to the amplitude range of the available experimental data. For the EXP-set, 4 thresholds were selected: 100\,mV, 200\,mV, 400\,mV and 600\,mV. The two crossing points of the threshold with the rising and falling edges of the waveform are used to integrate this region, where available, and calculate each $Q_{up}$ value of the corresponding pulse. To make available the information of the e-peak amplitude, read-out electronics to digitize the waveform
are required. However, on a large-scale experimental set-up, this will imply a significant number of channels, which can skyrocket the cost of the experimental set-up. Introducing pulse integrators such as the NINO-chip \cite{NINO}, or other new generation front-end electronics, in addition to the discriminator device, can achieve the time resolution of the Constant Fraction Discrimination Technique, at a significantly lower cost.

\begin{figure}[hbt!]
	\centering
	\includegraphics[width=0.4\linewidth]{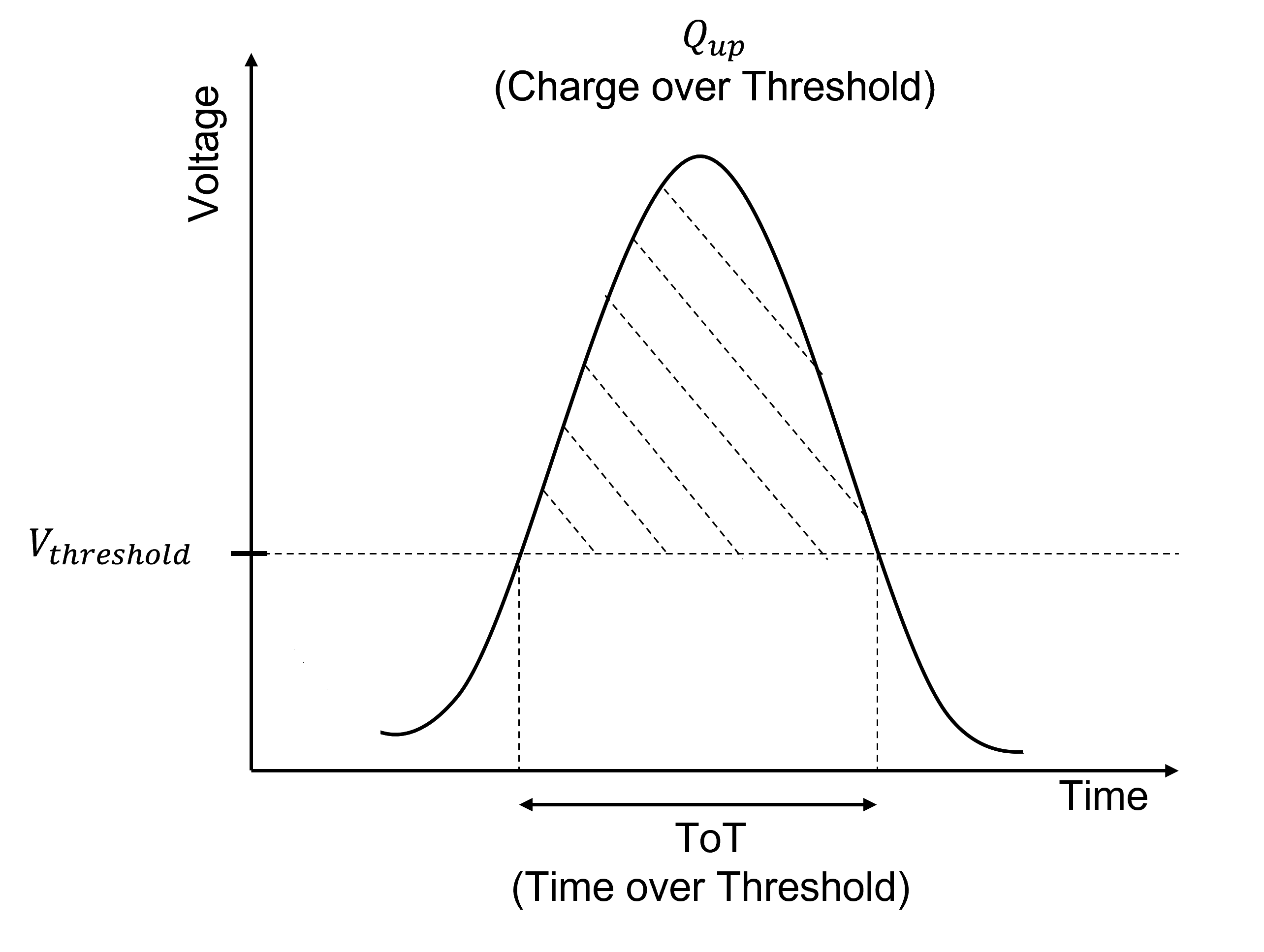}
	\caption{Definition of the charge above relevant threshold, $Q_{up}$. }
	\label{fig:qup_representation}
\end{figure} 
\begin{figure}[hbt!]
	\begin{subfigure}{0.5\textwidth}
		\centering 
		\includegraphics[width=0.7\textwidth]{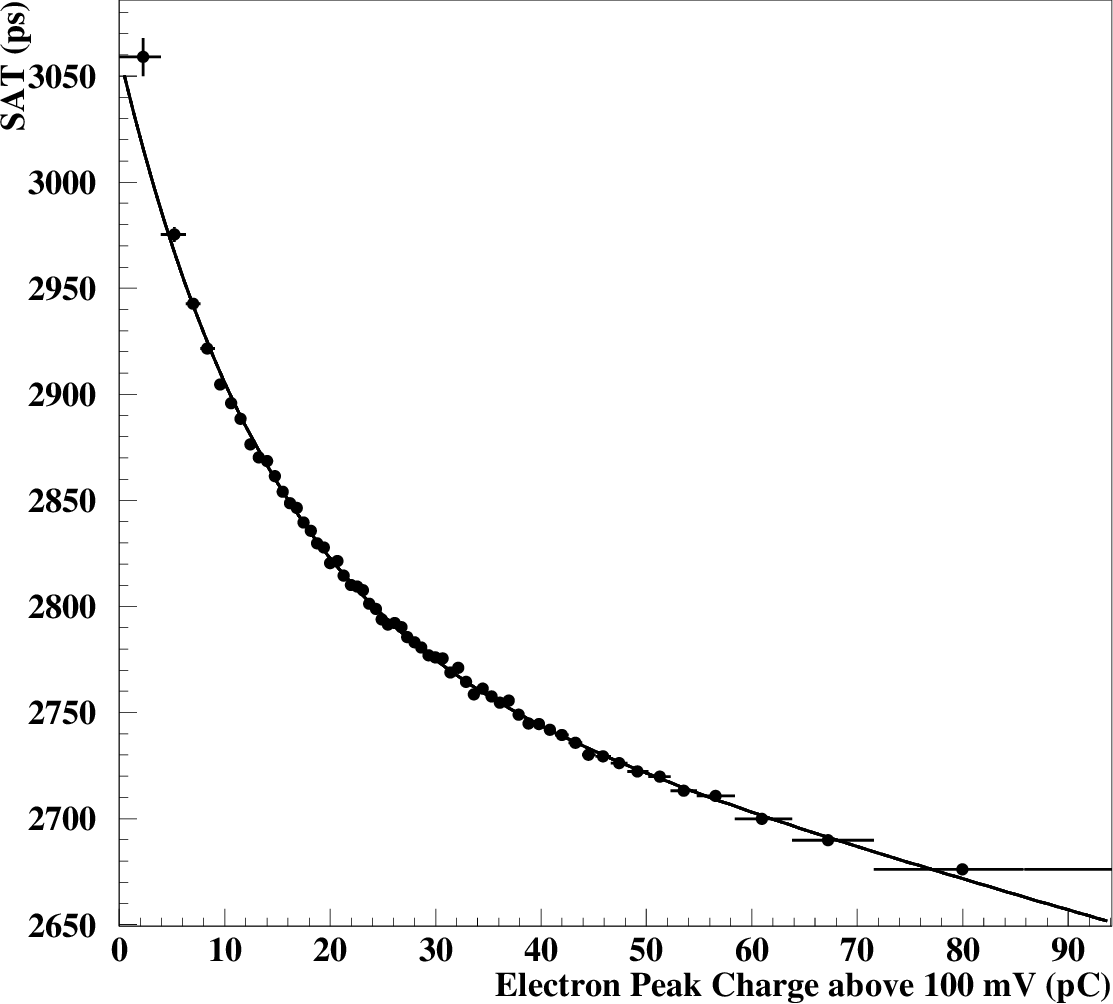}
		\caption{}\label{fig:qup1_slewing}
	\end{subfigure}
	\begin{subfigure}{0.5\textwidth}
		\centering 
		\includegraphics[width=0.7\textwidth]{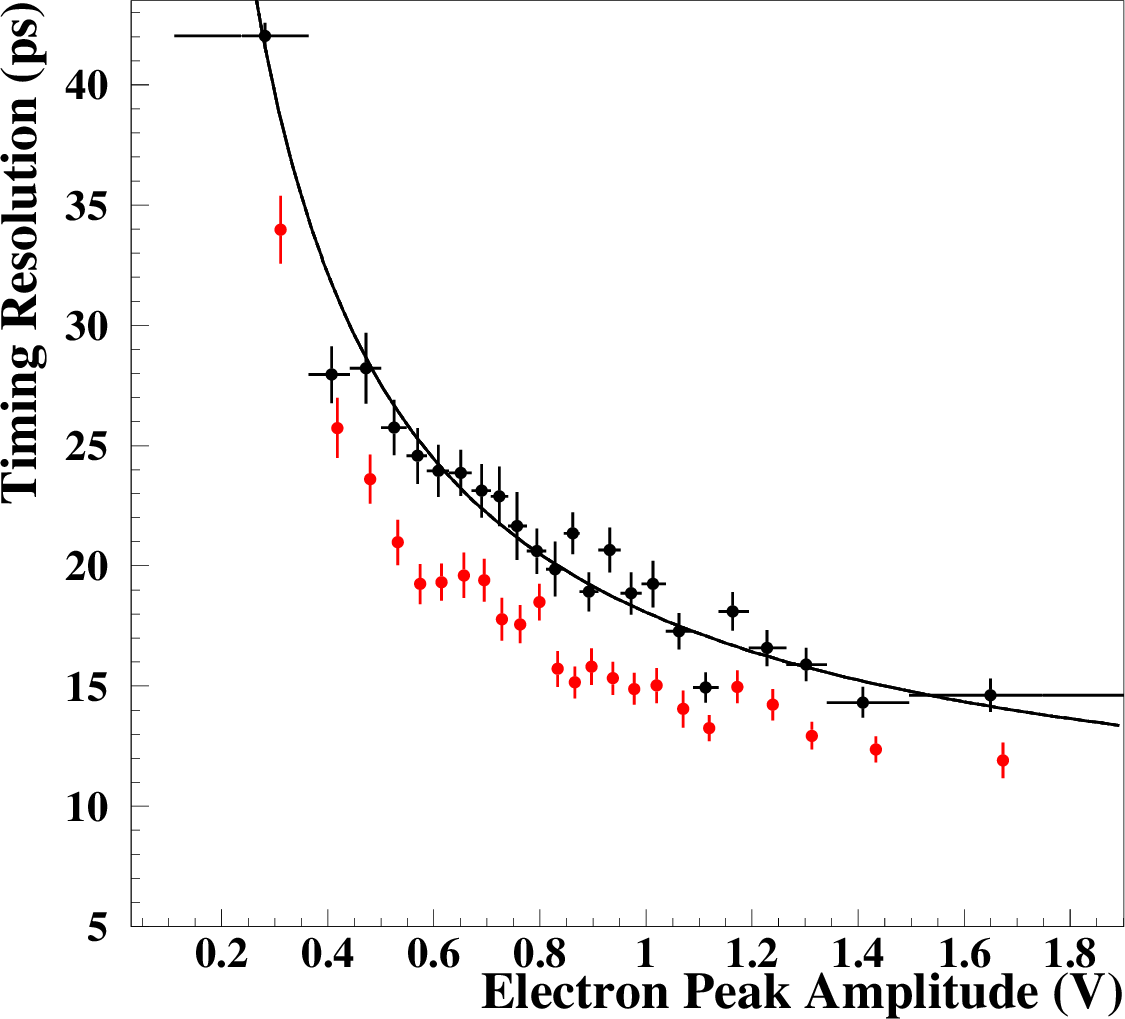}
		\caption{}\label{fig:corre_qup1}
	\end{subfigure}
	\caption{(a) SAT as a function of Charge above Threshold for the lowest threshold at 100\,mV, fitted with the corresponding power law curve. (b) Resolution as a function of e-peak amplitude after time walk correction, using both Constant Fraction Discrimination (red points) and multi-charge over threshold timing techniques (black points), for the same amplitude threshold.}\label{fig:qup_lower}
\end{figure}

\begin{figure}[hbt!]
	\begin{subfigure}{0.3\textwidth}
		\centering 
		\includegraphics[width=1.0\textwidth]{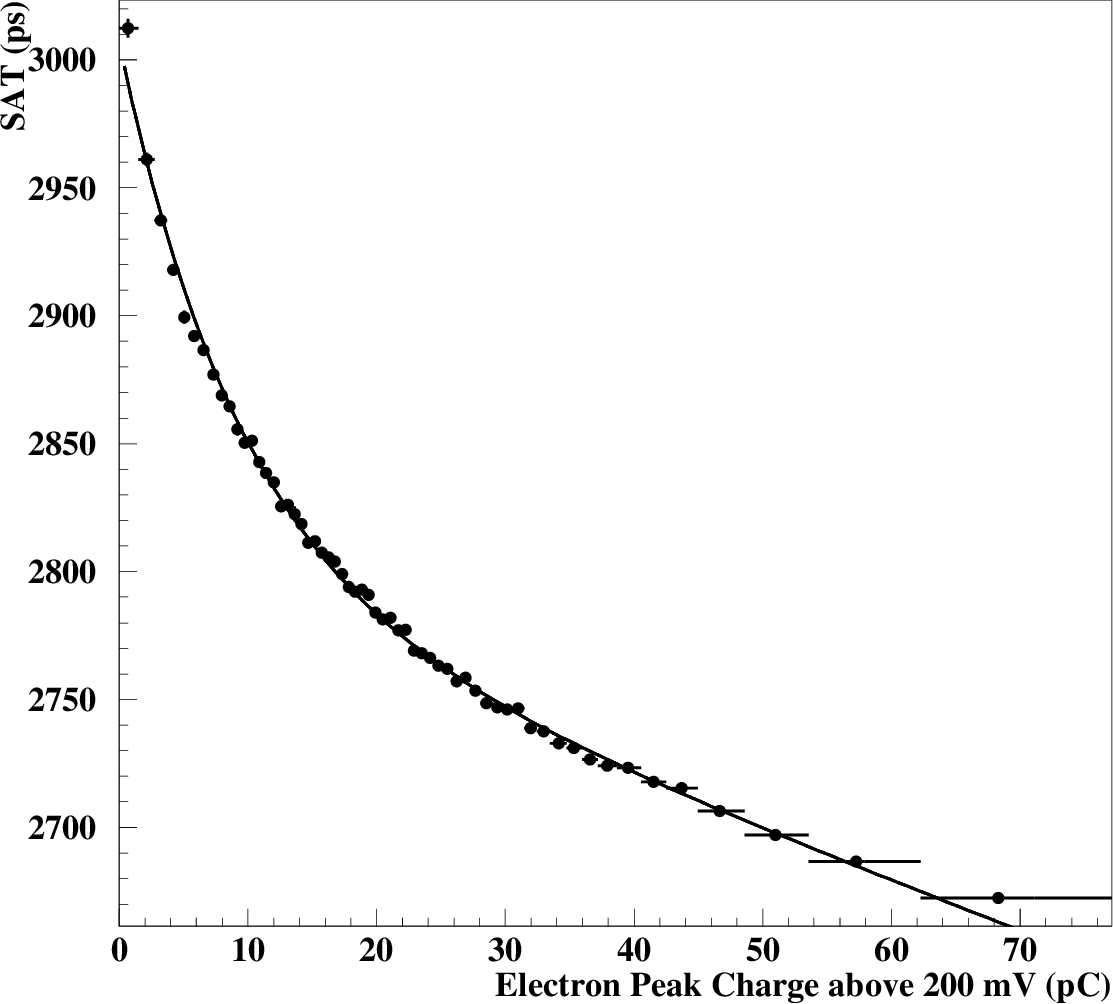}
		\caption{}\label{fig:qup2_slewing}
	\end{subfigure}
	\begin{subfigure}{0.3\textwidth}
		\centering 
		\includegraphics[width=1.0\textwidth]{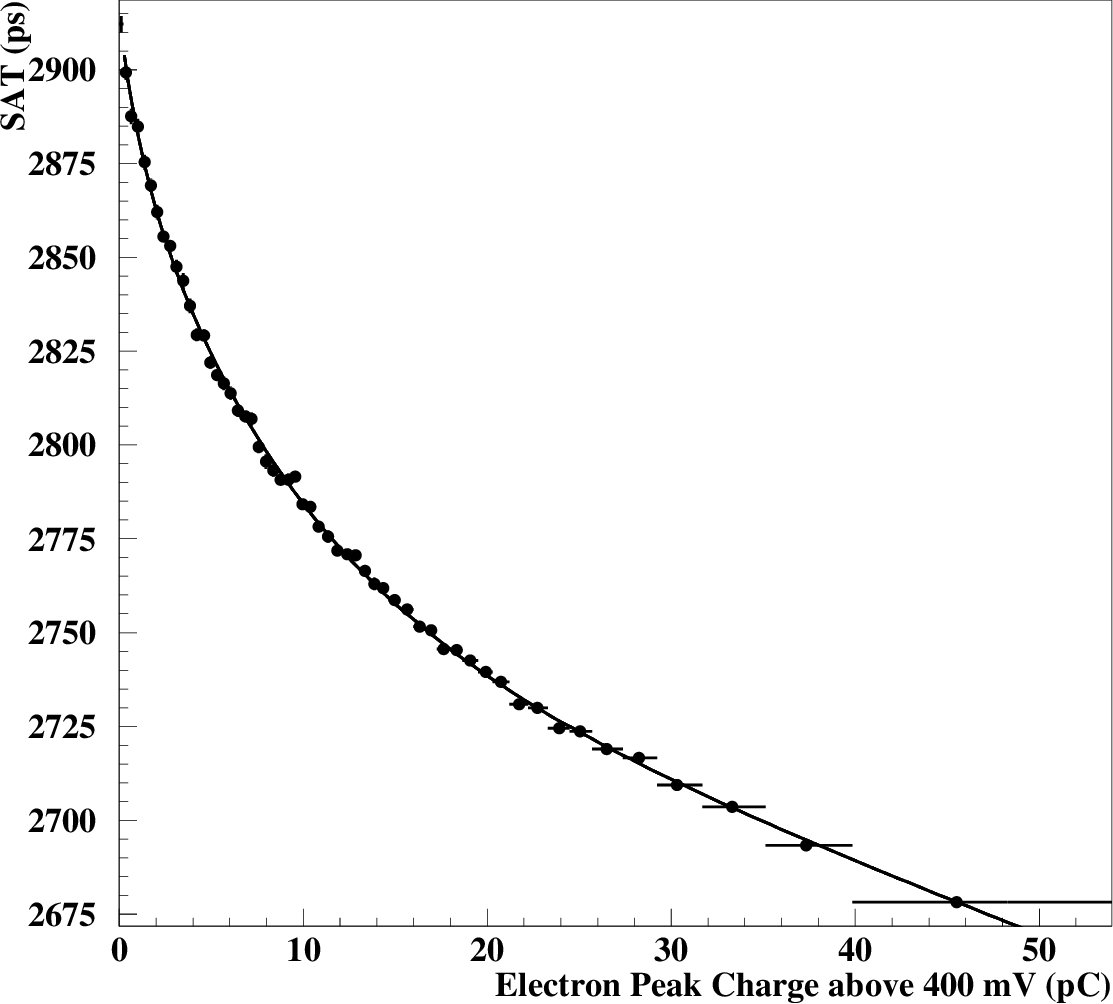}
		\caption{}\label{fig:qup3_slewing}
	\end{subfigure}
	\begin{subfigure}{0.3\textwidth}
		\centering 
		\includegraphics[width=1.0\textwidth]{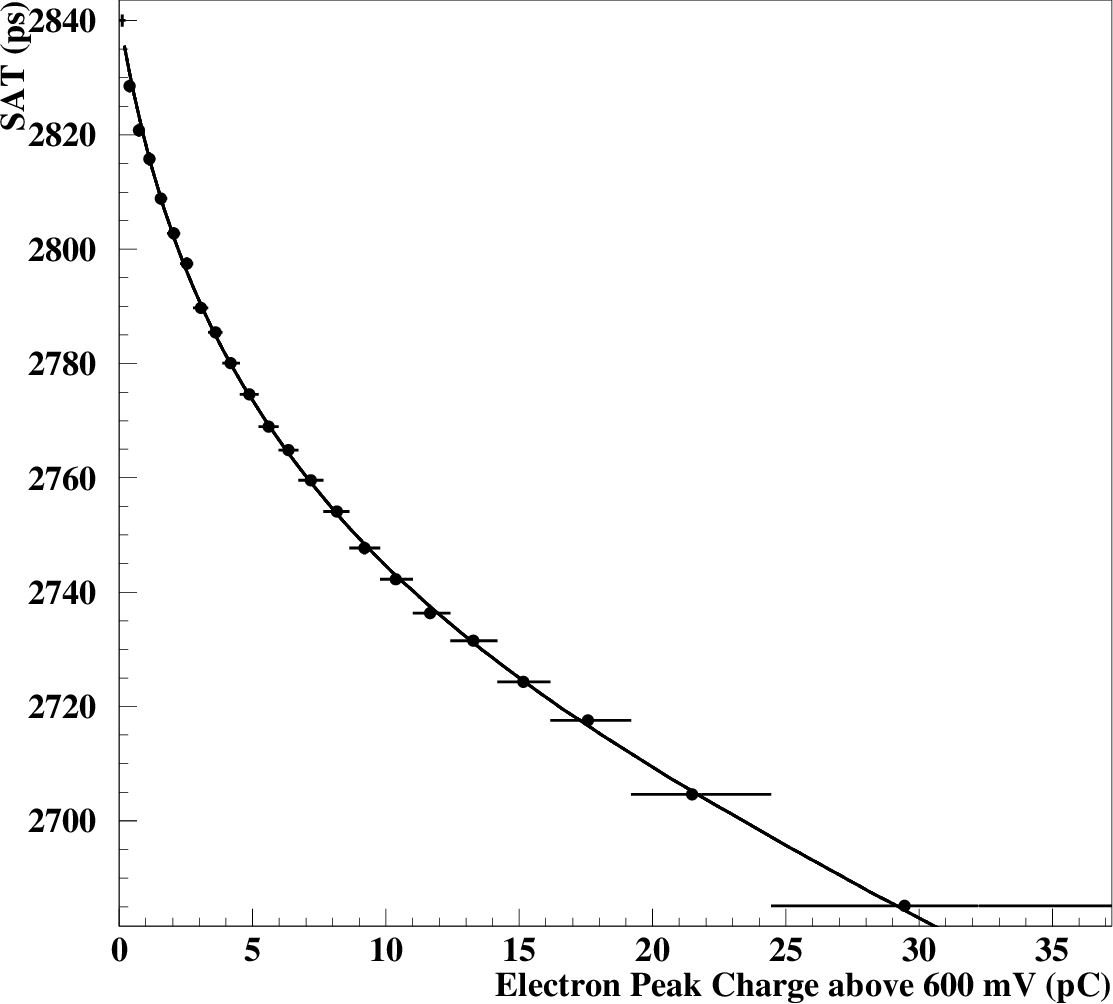}
		\caption{}\label{fig:qup4_slewing}
	\end{subfigure}
	\caption{SAT as a function of Charge above Threshold for the three other applied thresholds, fitted with the corresponding power law curve. Threshold at: (a) 200\,mV, (b) 400\,mV and (c) 600\,mV.}\label{fig:qup_slew_tot}
\end{figure}

The behavior of the mean SAT can be evaluated as a function of e-peak charge above the applied threshold (100\,mV), Fig.~\ref{fig:qup1_slewing}, parameterizing the time-walk effect using a power-law plus constant-term fit. Correcting the SAT for time walk using this fit is shown by the black points in Fig.~\ref{fig:corre_qup1}. The algorithm that provides the best time resolution uses the lowest threshold to define the SAT, and the parameterization as a function of $Q_{up}$ of the highest available threshold for the time-walk correction, on an event-by-event basis~\cite{kallitsopoulou2021development}. 
\added{Each pulse is corrected using the parameterization corresponding to the highest threshold crossed by that specific signal. Pulses that cross only one or two thresholds are still included, and no events are discarded from the dataset. This approach ensures that every waveform contributes to the final statistics while exploiting the most accurate correlation available for its amplitude range.}
This signal-processing algorithm minimizes the experimental information to a few numbers provided by the pulse integrators.


\begin{figure}[hbt!]
	\begin{subfigure}{0.3\textwidth}
		\centering 
		\includegraphics[width=1.0\textwidth]{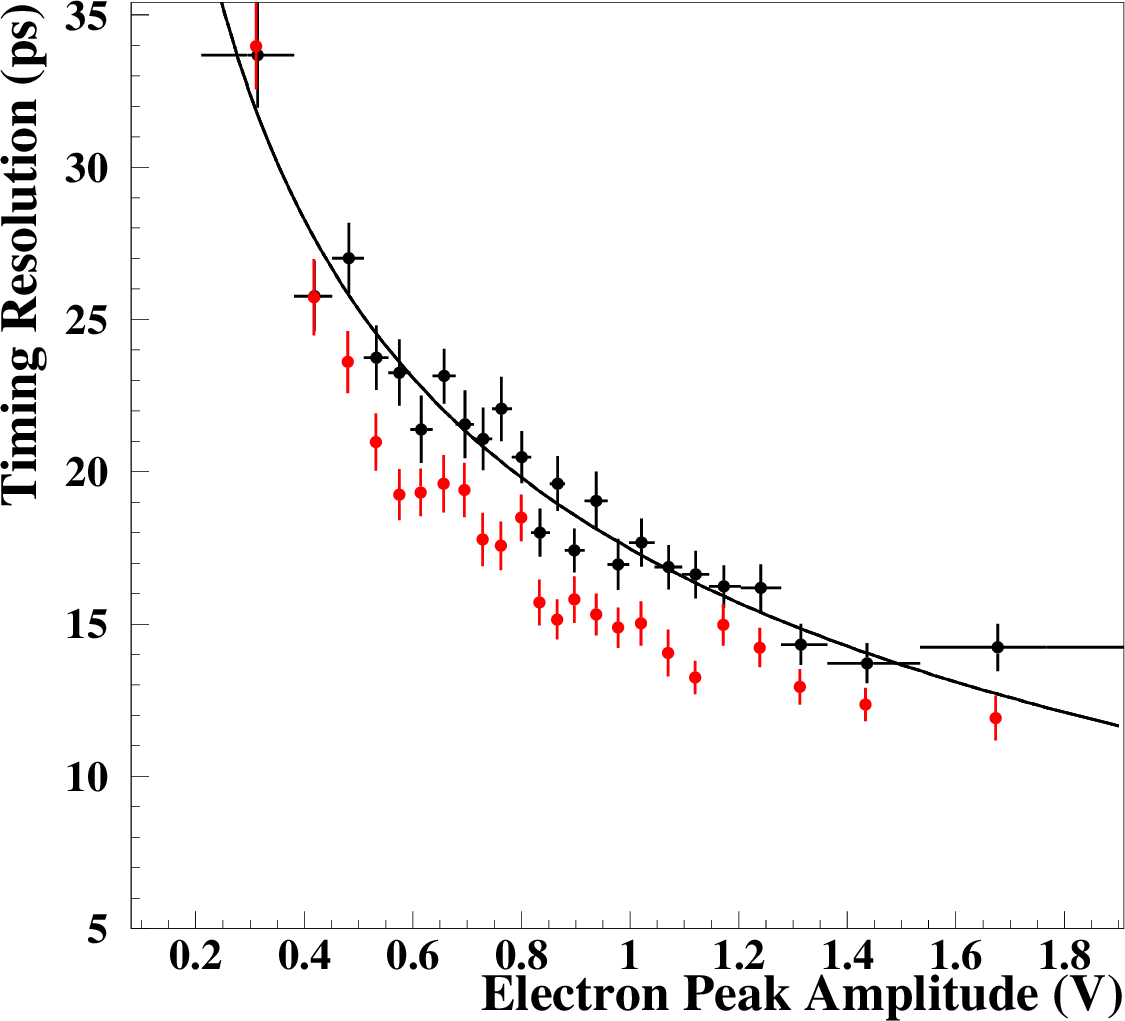}
		\caption{}\label{fig:corre_qup2}
	\end{subfigure}
	\begin{subfigure}{0.3\textwidth}
		\centering 
		\includegraphics[width=1.0\textwidth]{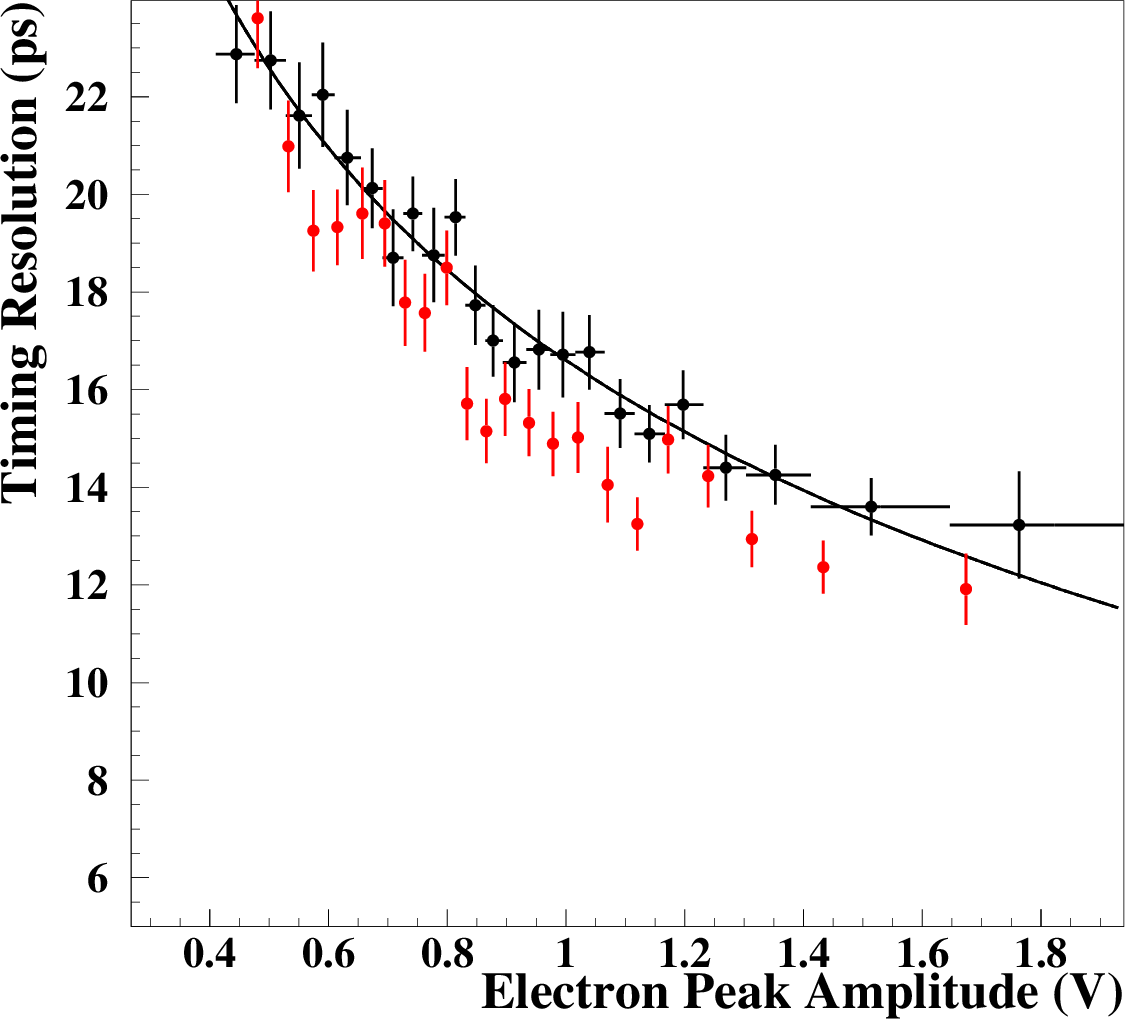}
		\caption{}\label{fig:corre_qup3}
	\end{subfigure}
	\begin{subfigure}{0.3\textwidth}
		\centering 
		\includegraphics[width=1.0\textwidth]{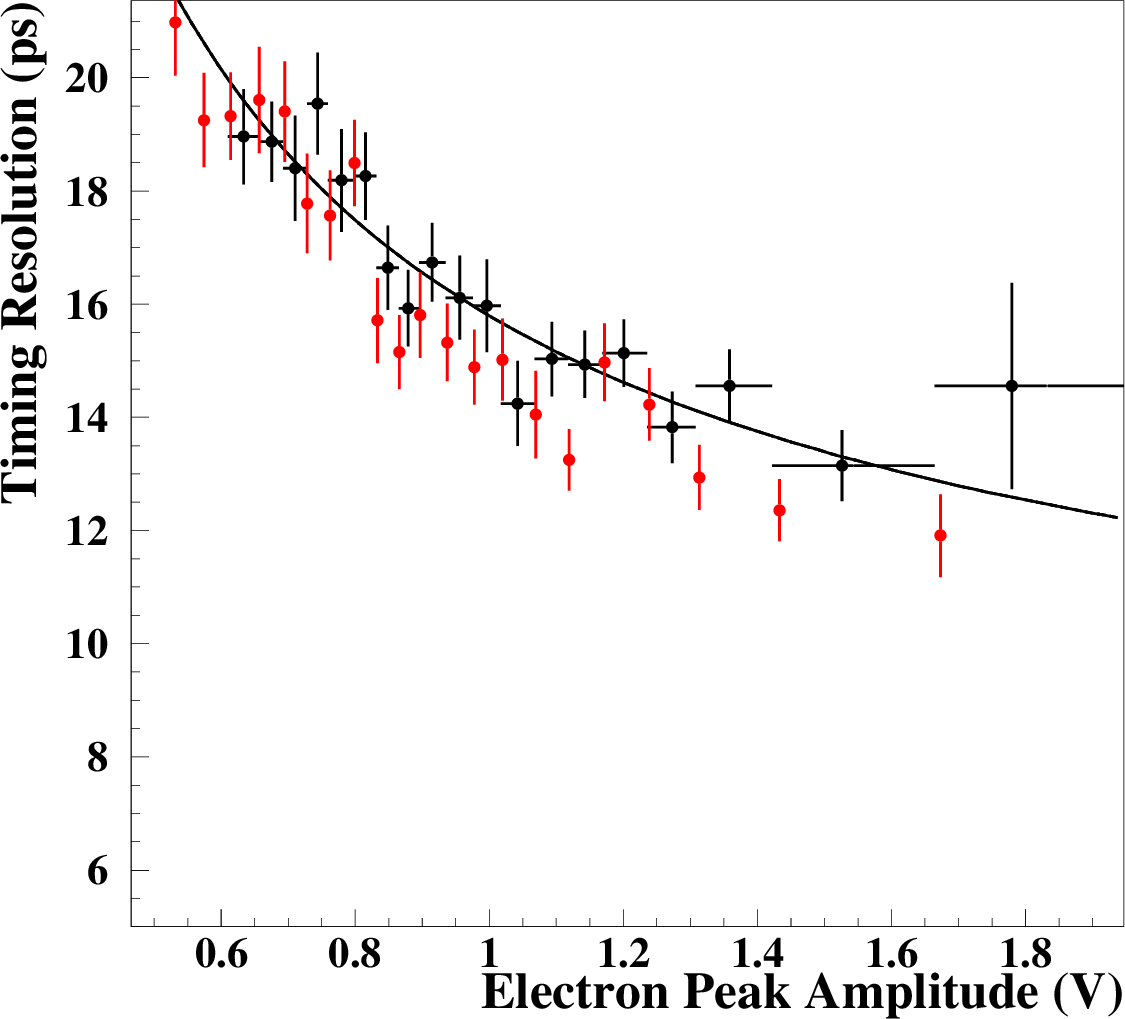}
		\caption{}\label{fig:corre_qup4}
	\end{subfigure}
	\caption{Time resolution as a function of e-peak amplitude after time walk correction, using both Constant Fraction Discrimination (red) and multi-Charge over threshold timing techniques (black), for threshold at (a) 200\,mV, (b) 400\,mV, (c) 600\,mV.}\label{fig:corre_qup_tot}
\end{figure}
\begin{figure}[hbt!]
	\begin{subfigure}{0.5\textwidth}
		\centering
		\includegraphics[width=0.8\linewidth]{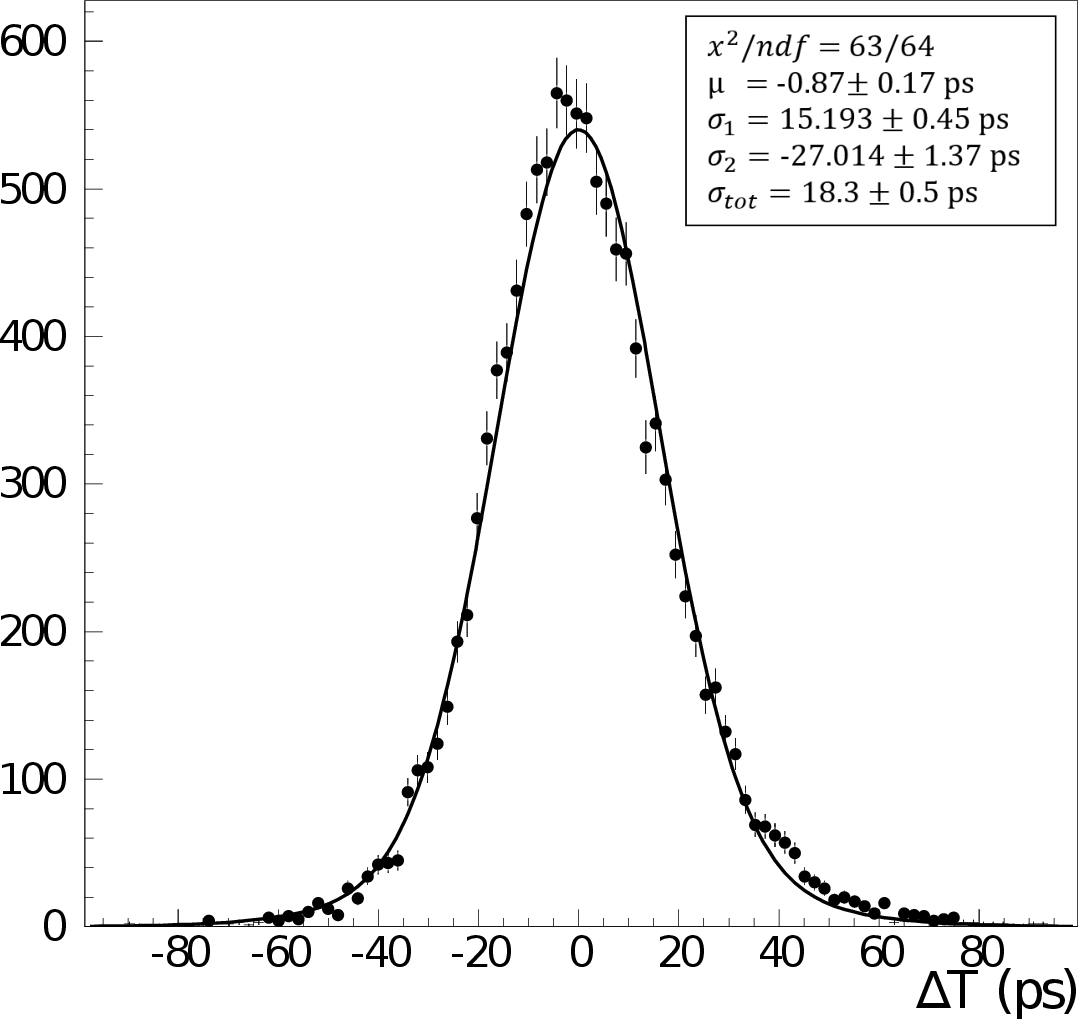}
		\caption{}
		\label{fig:pull_corre_ToT}
		
	\end{subfigure}
	\begin{subfigure}{0.5\textwidth}
		\centering
		\includegraphics[width=0.8\linewidth]{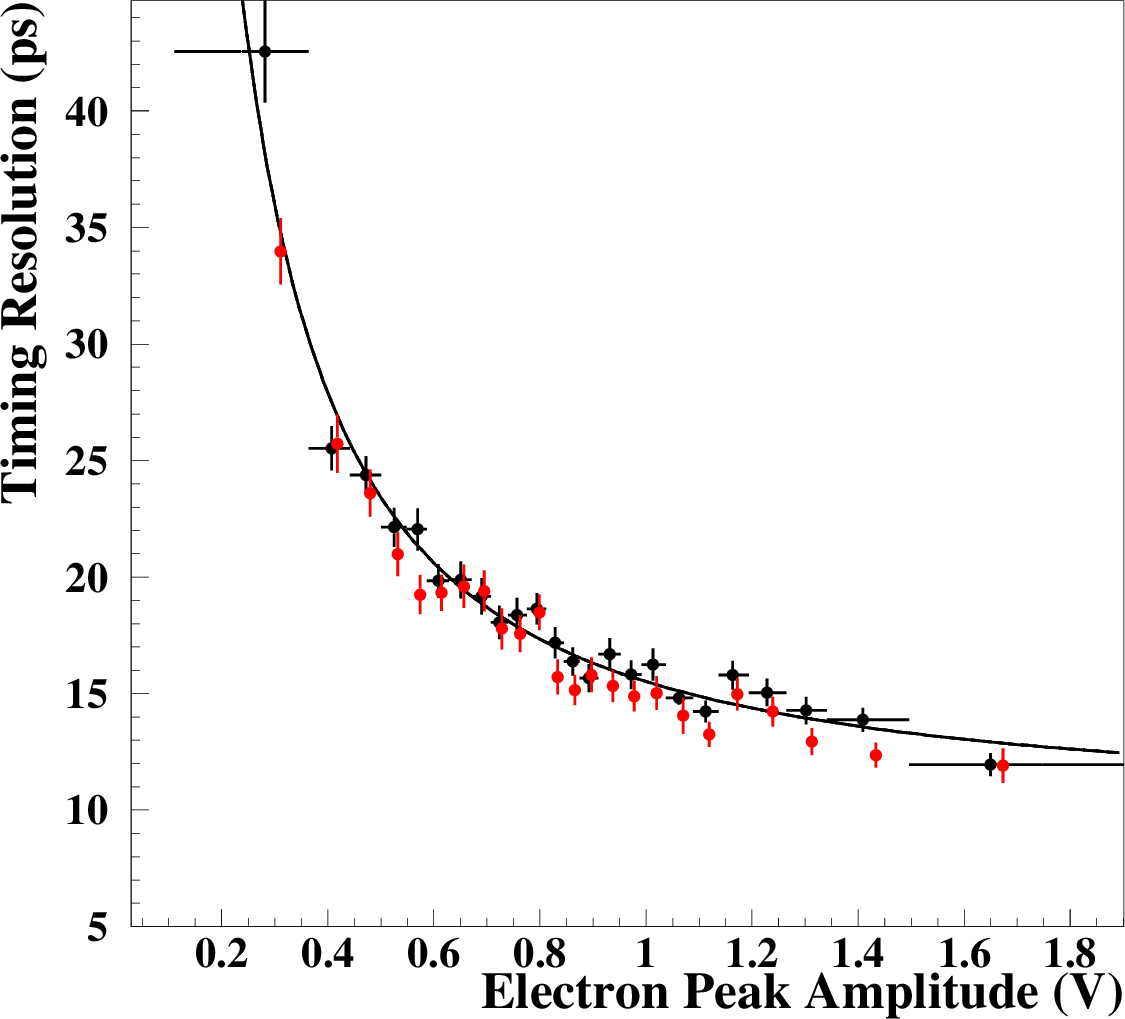}
		\caption{}
		\label{fig:corre_global}
	\end{subfigure}
	\caption{(a) SAT distribution of EXP-set after time walk corrections, using the multi-Charge over threshold Technique, resulting in a global timing resolution of 18.3$\pm$0.5\,ps.(b) Resolution as a function of e-peak amplitude after time walk correction, using Constant Fraction Discrimination (red points) and multi-charge-over-threshold timing techniques (black points).}
\end{figure}

The event-by-event time walk correction using the calibration curve of the higher available crossing threshold requires calibration measurements before any analysis ( Fig.\ref{fig:qup_slew_tot} ). The analysis of the EXP-set, using the $Q_{up}$ technique, is shown in Fig.\ref{fig:pull_corre_ToT}, resulting in a timing resolution performance of the detector, as the one presented in Fig. \ref{fig:corre_global}, with black points. 
This is similar to the one achieved with the CFD Technique, which is expressed with red points in the same Figure and signifies the successful performance of the timing algorithm. After correction for systematic effects, SAT distribution is symmetric, with an RMS of 18.3$\pm$0.5\,ps. The same results were achieved with the full offline analysis with CFD reported also in \cite{kallitsopoulou2021development}.

\section{An Emulation Model for PICOSEC-MM Pulses}\label{sec:simulation}

\subsection{Characterization of Single-Photoelectron Waveforms Properties}
While analytical signal processing algorithms provide valuable benchmarks for timing extraction, their complexity can limit applicability in real-time environments. To address this, previous studies, such as \cite{Manthos_2021, Tsiamis_2022}, have shown that ANN techniques can be used for online estimation of the arrival time of PICOSEC-MM waveforms with accuracy similar to the results of the full offline analysis. However, these studies were limited by the size of the available ANN training set and therefore relied on k-fold cross-validation methods for analysis \cite{francois}. In this work, a method for simulating large sets of waveforms is developed to train the ANN and provide more conclusive results.

\begin{figure}[hbt!]
	\centering 
	\includegraphics[width=0.4\textwidth]{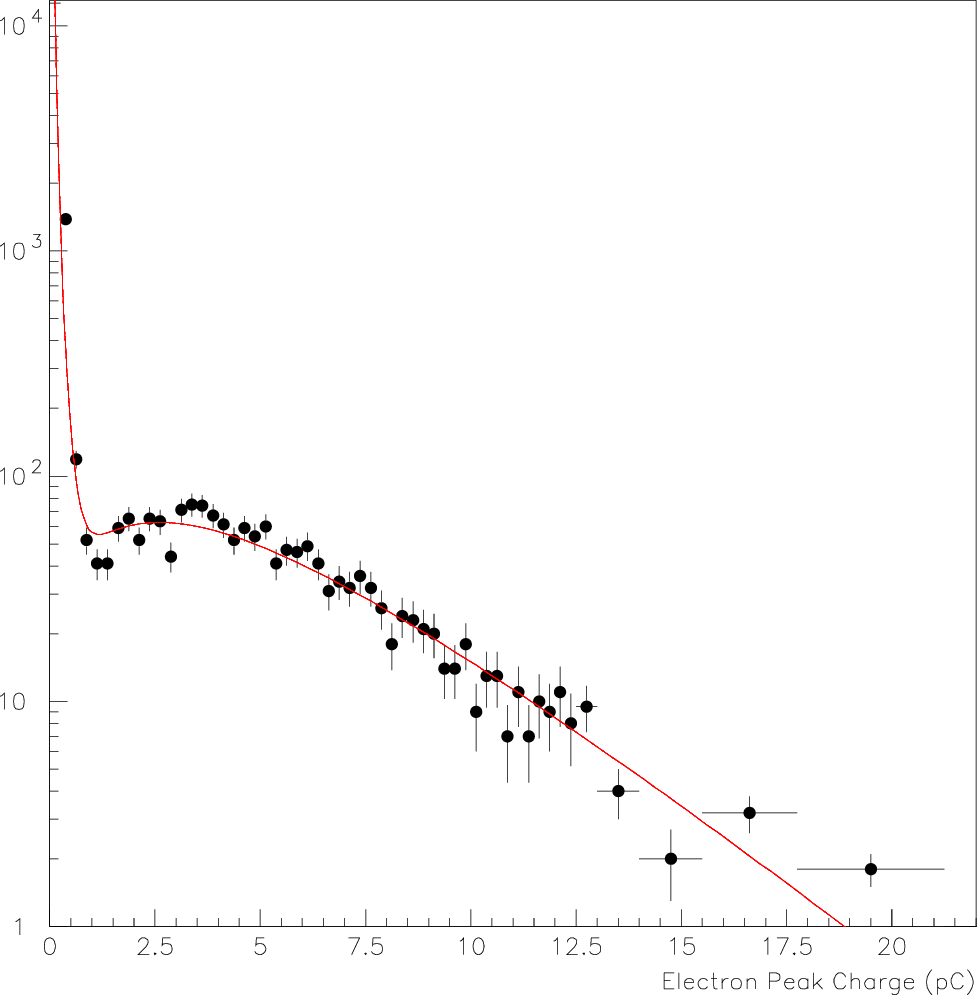}
	\caption{Single Photo-electron Charge Distribution, of the SPE-set. The red solid line corresponds to the fitting curve, the convolution of an exponential for the falling edge of the noise peak, and a Polya distribution for the rest of the data.}
	\label{fig:singlepe_slew}
\end{figure}

The fully digitized waveforms of the SPE-set are used to emulate waveforms of the EXP-set, on the response of the PICOSEC-MM to a fewphotoelectrons. As it has been previously shown (e.g.\cite{BORTFELDT2018317, BORTFELDT2021165049}), the e-peak charge distribution of the single-pe waveform is well described by the Gamma probability distribution function (pdf) of Eq.\ref{eq:1-polya}:  

\begin{equation}\label{eq:1-polya}
	P_1(Q;\bar{Q},\theta)  = \frac{(1+\theta)^{(1+\theta)}}{\bar{Q} \Gamma(1+\theta)}\Big(\frac{Q}{\bar{Q}}\Big)^{\theta} exp\Big[-(1+\theta)\frac{Q}{\bar{Q}}\Big]
\end{equation}
where $\bar{Q}$ is the mean value of the distribution and $\alpha=\theta + 1$ is the shape parameter. Fig.\ref{fig:singlepe_slew} shows the e-peak charge distribution of the SPE-set while the solid line is the graphical representation of Eq.\ref{eq:1-polya} with $\bar{Q} = 0.126$\,pC and $\theta = 1.72$ (corresponding to an RMS of 0.077 \,pC). 

\begin{figure}[hbt!]
	\centering
    \includegraphics[width=0.5\textwidth]{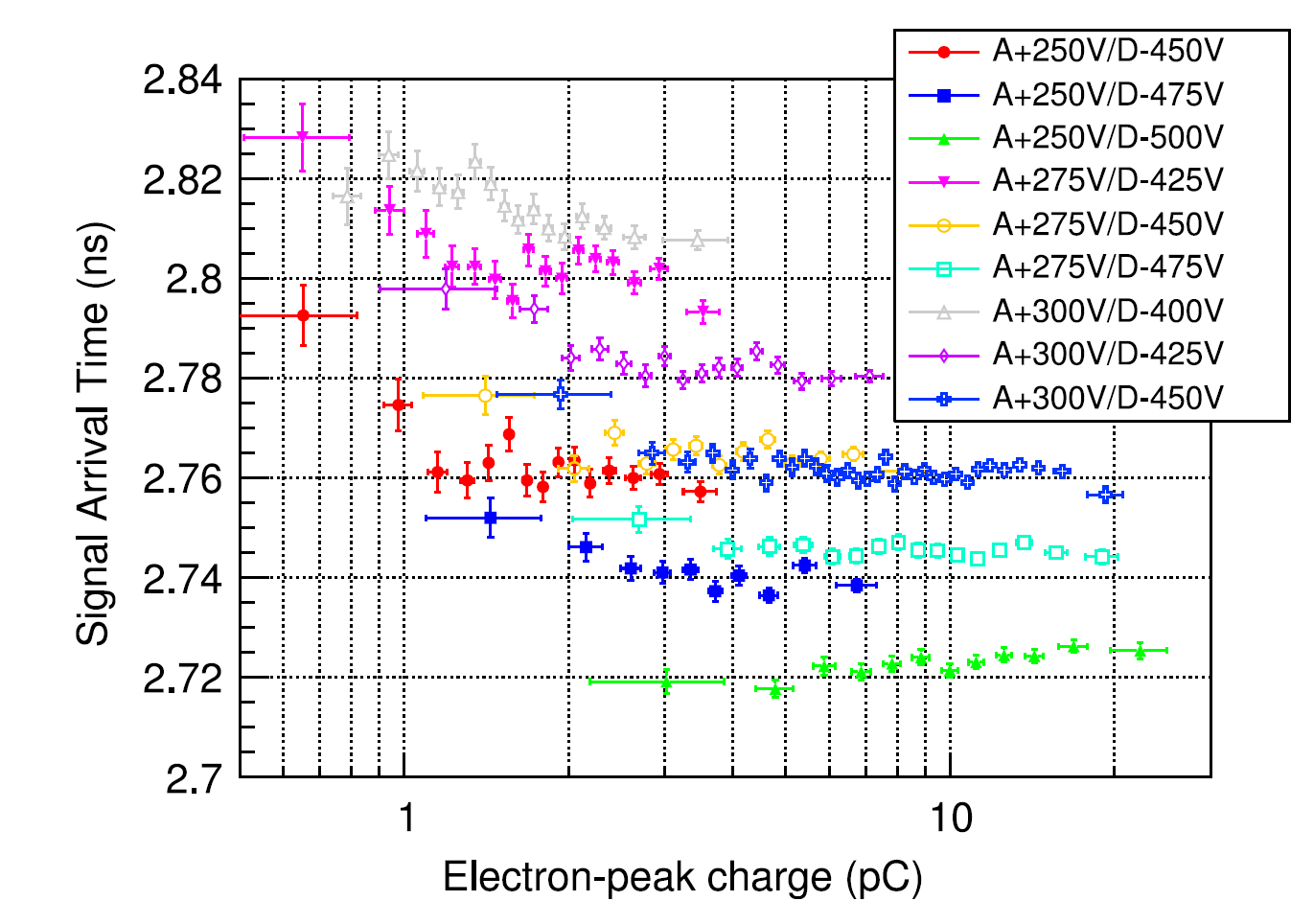}
   \caption{\added{Dependence of the SAT on the electron-peak charge for different anode–drift (A–D) voltage configurations, shown in different colors~\cite{BORTFELDT2021165049}.}}
   \label{fig:sat_high_gain}
\end{figure}

\begin{figure}[hbt!]
	\begin{subfigure}{0.5\textwidth}
		\centering 
		\includegraphics[width=0.8\linewidth]{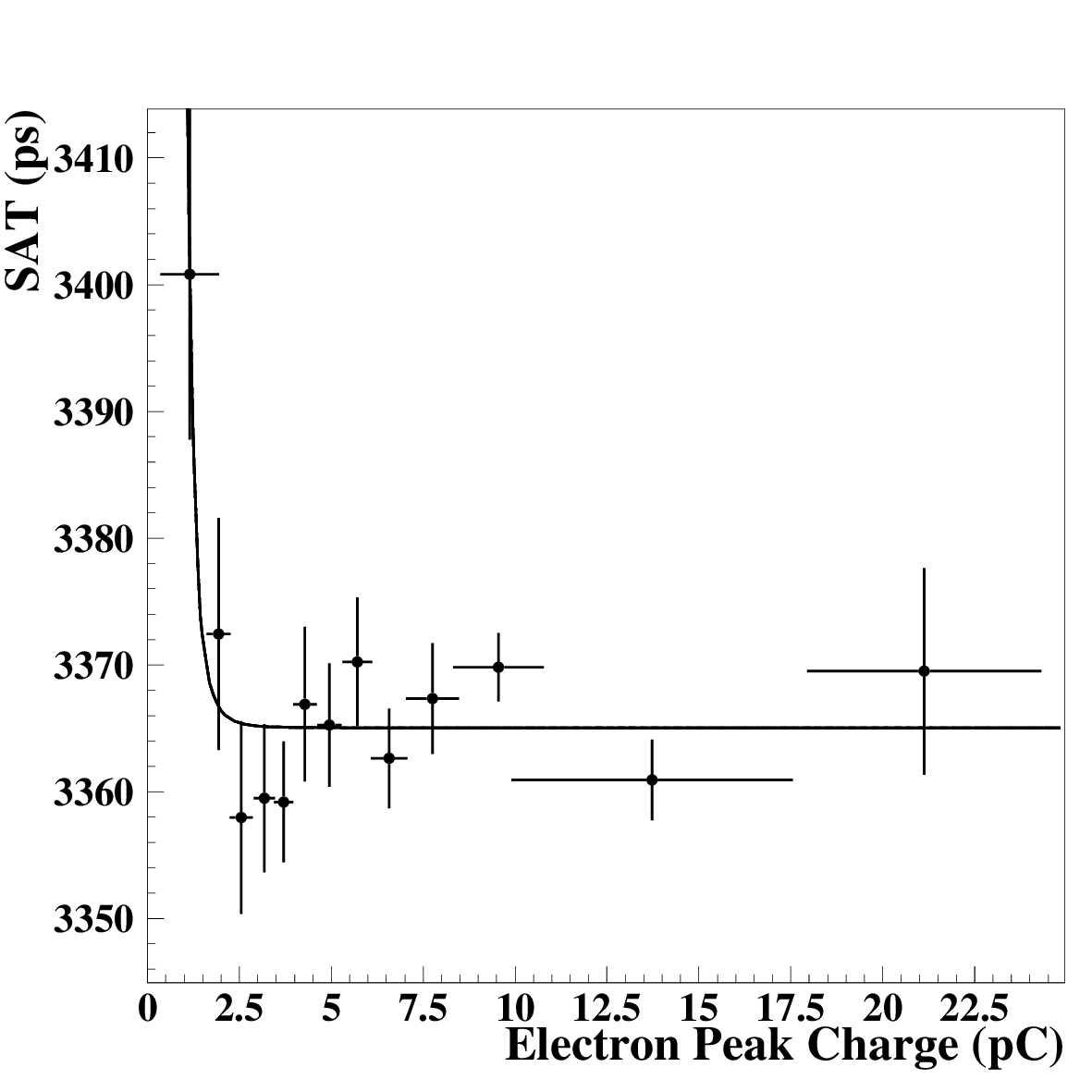}
		\caption{}
		\label{fig:slew_spe}
	\end{subfigure}
	\begin{subfigure}{0.5\textwidth}
		\centering 
		\includegraphics[width=0.8\textwidth]{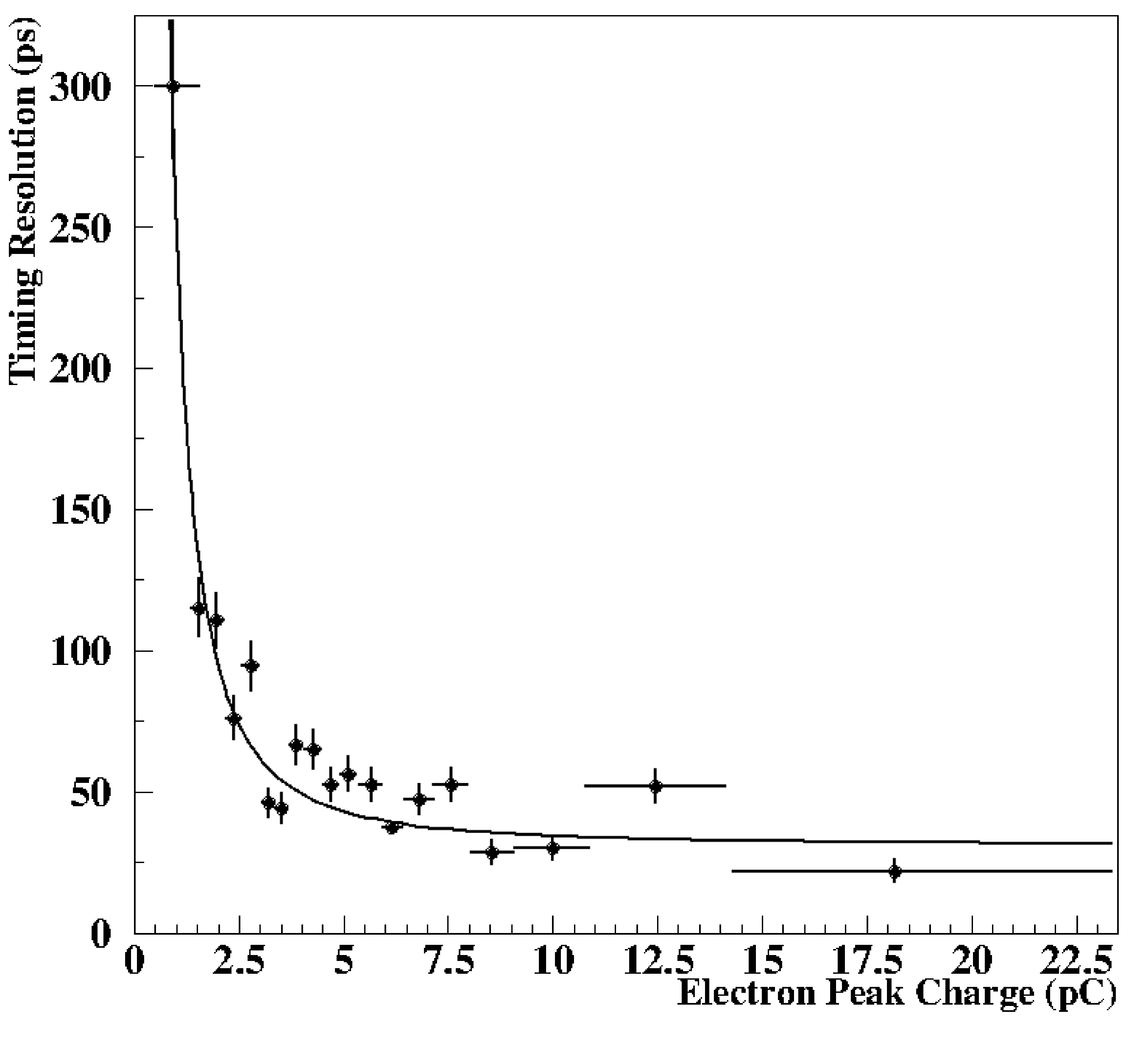}
		\caption{}
		\label{fig:single1}
	\end{subfigure}
	\caption{(a) Dependence of the SAT as a function of e-peak charge, for the SPE-set pulses. The solid line represents a power law fit to the experimental data. (b) Timing resolution as a function of the e-peak charge of the SPE-set. The solid line represents a double exponential fit to the experimental data.}
\label{fig:SPE-all}
\end{figure}

\begin{figure}[hbt!]
	\begin{subfigure}{0.5\textwidth}
	\centering
	\includegraphics[width=0.8\textwidth]{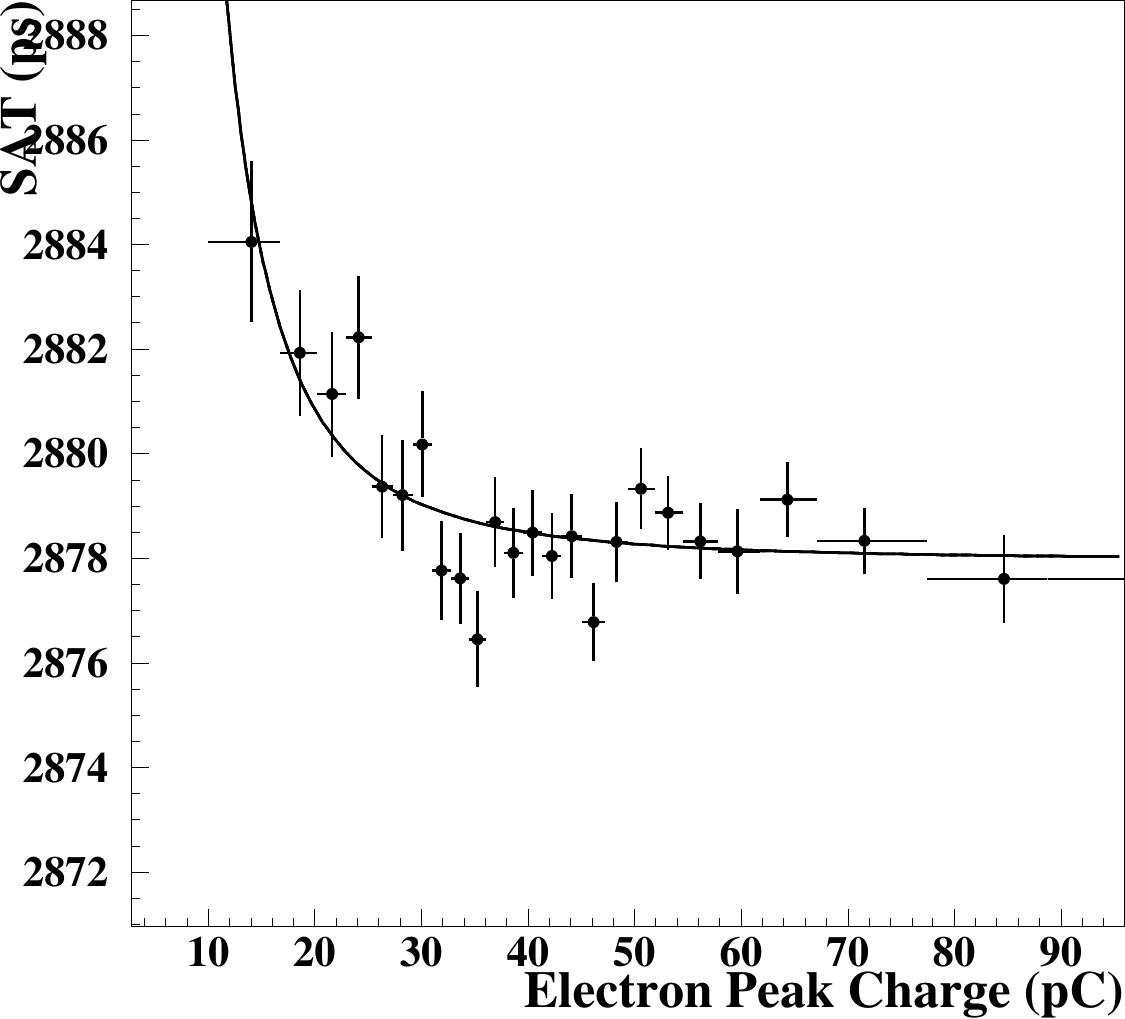}
	\caption{}
	\label{fig:slew_exp}
	\end{subfigure}
	\begin{subfigure}{0.5\textwidth}
	\centering 
	\includegraphics[width=0.8\textwidth]{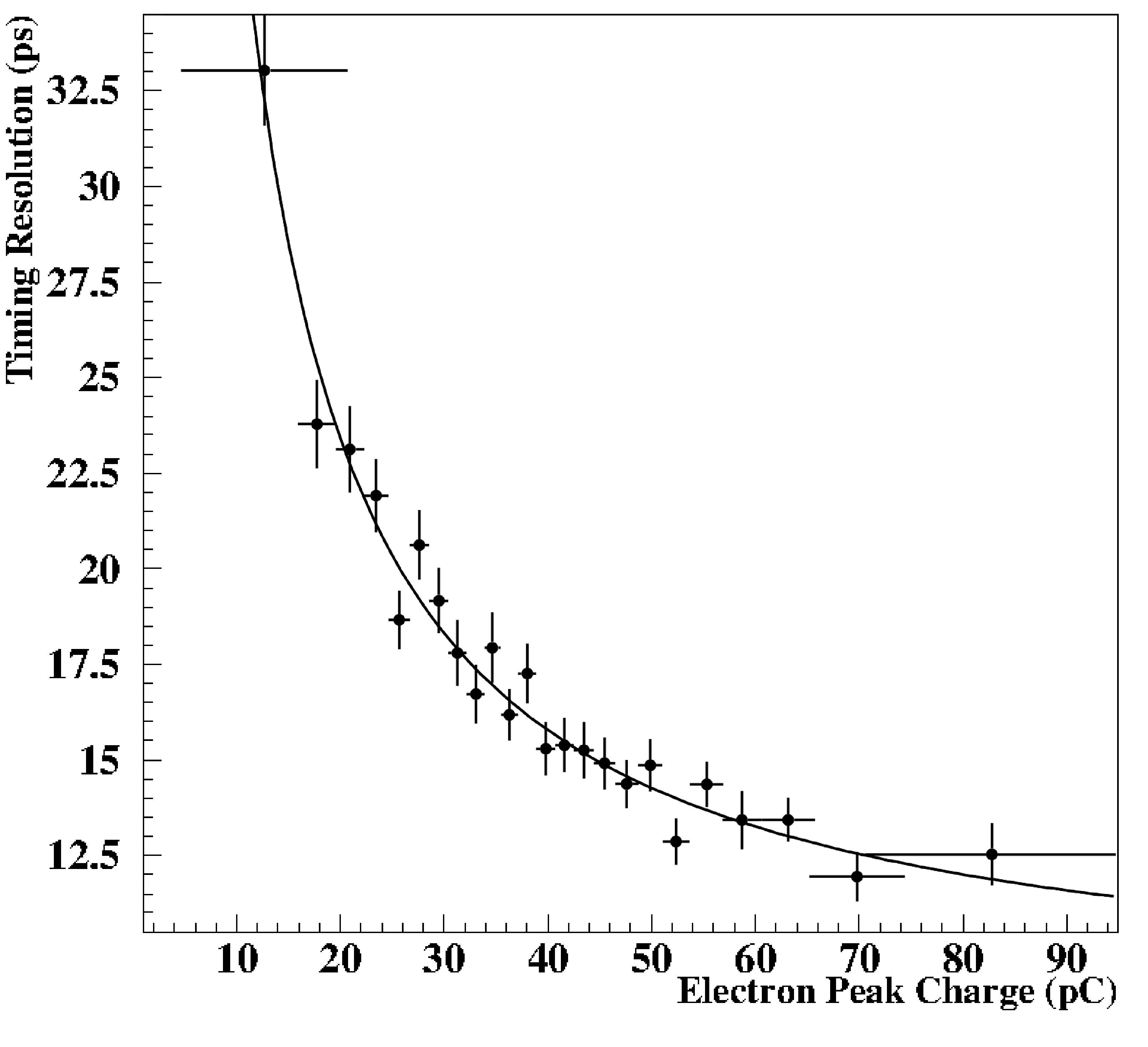}
	\caption{}
	\label{fig:single2}
	\end{subfigure}
	\caption{(a) Dependence of the SAT as a function of e-peak charge, for the EXP-set pulses.(b) Timing resolution as a function of e-peak charge for the EXP-set. The solid line corresponds to a power law fit to the data points, both for the SAT and Timing Resolution data points.}
 \label{fig:EXP-all}
\end{figure}

Operating the detector in such a high gain (i.e. 44\,$\mathrm{kV/cm}$ at the drift and over 21\,$\mathrm{kV/cm}$ at the amplification stage, generated by 525/275\,V ), while creating the two data-sets, leads to a very small dependence of the pulses arrival time on their size, meaning that there is almost no time walk effect, as can be seen in Fig.\ref{fig:sat_high_gain}. \added{Although the voltage settings used for the data presented in this work are not included here, the figure illustrates the general trend that higher electric fields lead to a reduced time-walk effect.}
Using the full offline analysis algorithm, we can extract the timing performance of the two data sets as a function of the e-peak charge. Fig.\ref{fig:slew_spe}, shows the SAT as a function of the e-peak charge, confirming the small effect of the time-walk on the single photoelectron data-set, and it's corresponding resolution, as in Fig.\ref{fig:single1}. Respectively, Fig.\ref{fig:EXP-all} represents the SAT of the EXP-set and its corresponding resolution, as a function of the e-peak charge. 

\subsection{Synchronizing and Summing Single Photo-electron Waveforms}


In the case that the PICOSEC-MM signal is produced by $N_{pe}$ photoelectrons the charge distribution of the corresponding e-peak is described by the convolution of $N_{pe}$ identical pdfs, following the functional form of  Eq.\ref{eq:1-polya}. This convolution results in the multi-pe pdf of Eq.\ref{eq:N-polya}: 

\begin{equation}\label{eq:N-polya}
	P_N(Q;\bar{Q},\theta,N_{pe})  = \frac{(1+\theta)^{N_{pe}(1+\theta)}}{\bar{Q} \Gamma(N_{pe}(1+\theta))}\Big(\frac{Q}{\bar{Q}}\Big)^{N_{pe}(\theta+1)-1} exp\Big[-(1+\theta)\frac{Q}{\bar{Q}}\Big]
\end{equation}


\begin{figure}[hbt!]
	
		\centering
		\includegraphics[width=0.5\linewidth]{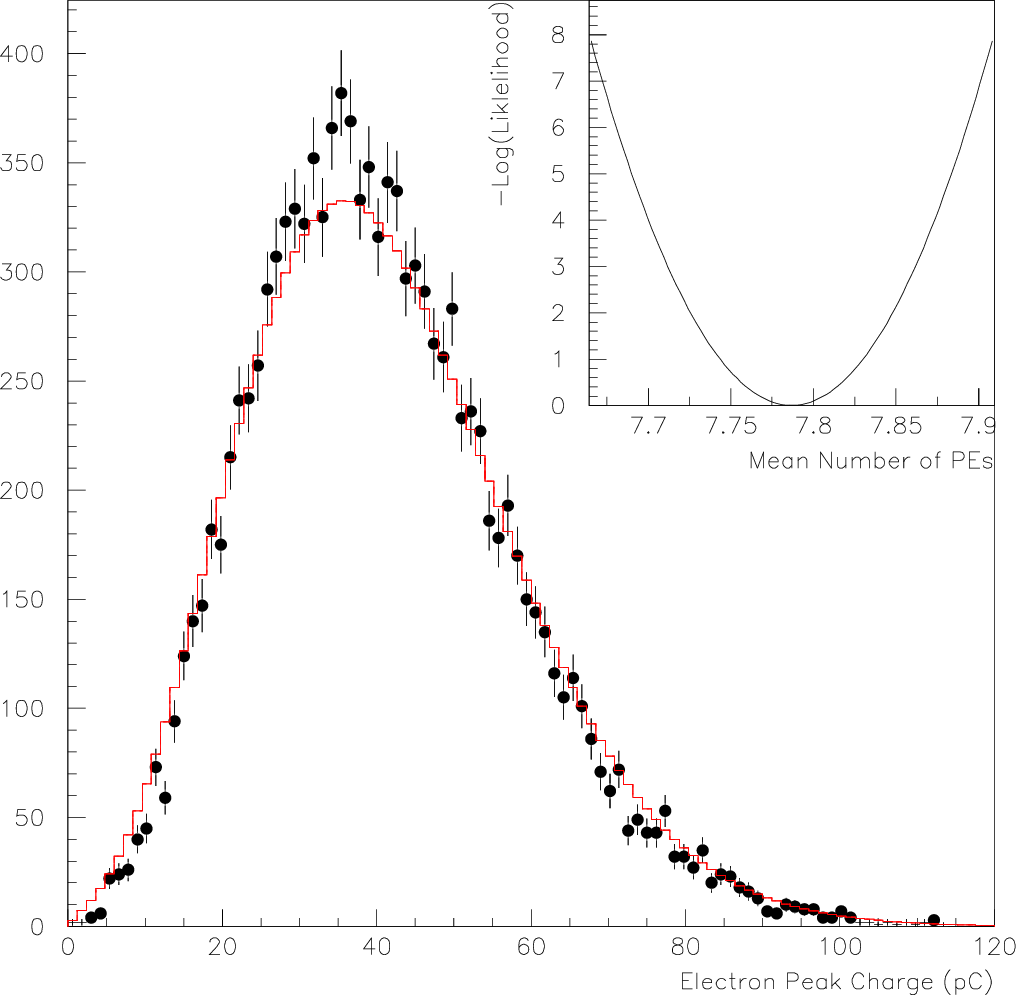}
		\caption{(a)E-peak Charge Distribution of EXP-set. With red solid line is the prediction of the deconvolution algorithm, as a fit to the experimental data (black points). In the inset plot the Log-Likelihood minimization for the estimation of the number of photoelectrons.}
		\label{fig:deconvolution}

\end{figure} 

As in \cite{Manthos_2020}, we assume that the EXP-set comprises PICOSEC-MM waveforms corresponding to $N_{pe}$ photoelectrons, where $N_{pe}$ follows a Poisson distribution with a mean value equal to $\mu_{pe}$. Then, the e-peak charge distribution of the EXP-set waveforms should follow the following pdf: 

\begin{equation}\label{eq:convol}
	G(Q; \mu_{pe}, \bar{Q}, \theta) = \sum_{N_{pe}=0}^{\infty} \frac{\mu_{pe}^{N_{pe}}\cdot e^{-\mu_{pe}}}{N_{pe}!} \cdot	P_N(Q;\bar{Q},\theta,N_{pe})
\end{equation}
where $P_N(Q;\bar{Q},\theta,N_{pe})$ is given by Eq.\ref{eq:N-polya}.

The e-peak charge distribution of the EXP-set waveforms is represented with the solid red line of Fig.\ref{fig:deconvolution}. The mean value of the number of photoelectrons, that have produced these waveforms, has been estimated by fitting Eq.\ref{eq:convol} to the above data points. In this fit, the $\bar{Q}$ and $\theta$ parameter values have been fixed to  0.126\,pC and 1.72 respectively, which describe well the single-pe charge distribution (Fig.\ref{fig:singlepe_slew}). The fit results to $\hat{\mu}_{pe} = $ 7.8$\pm$0.1 \,pes. A graphical representation of Eq.\ref{eq:convol} with $\hat{\mu}_{pe} = 7.8$ is shown, as the solid red curve, in Fig.\ref{fig:deconvolution}.

\begin{figure}[hbt!]
	\begin{subfigure}{0.5\textwidth}
		\centering
		\includegraphics[width=0.7\linewidth]{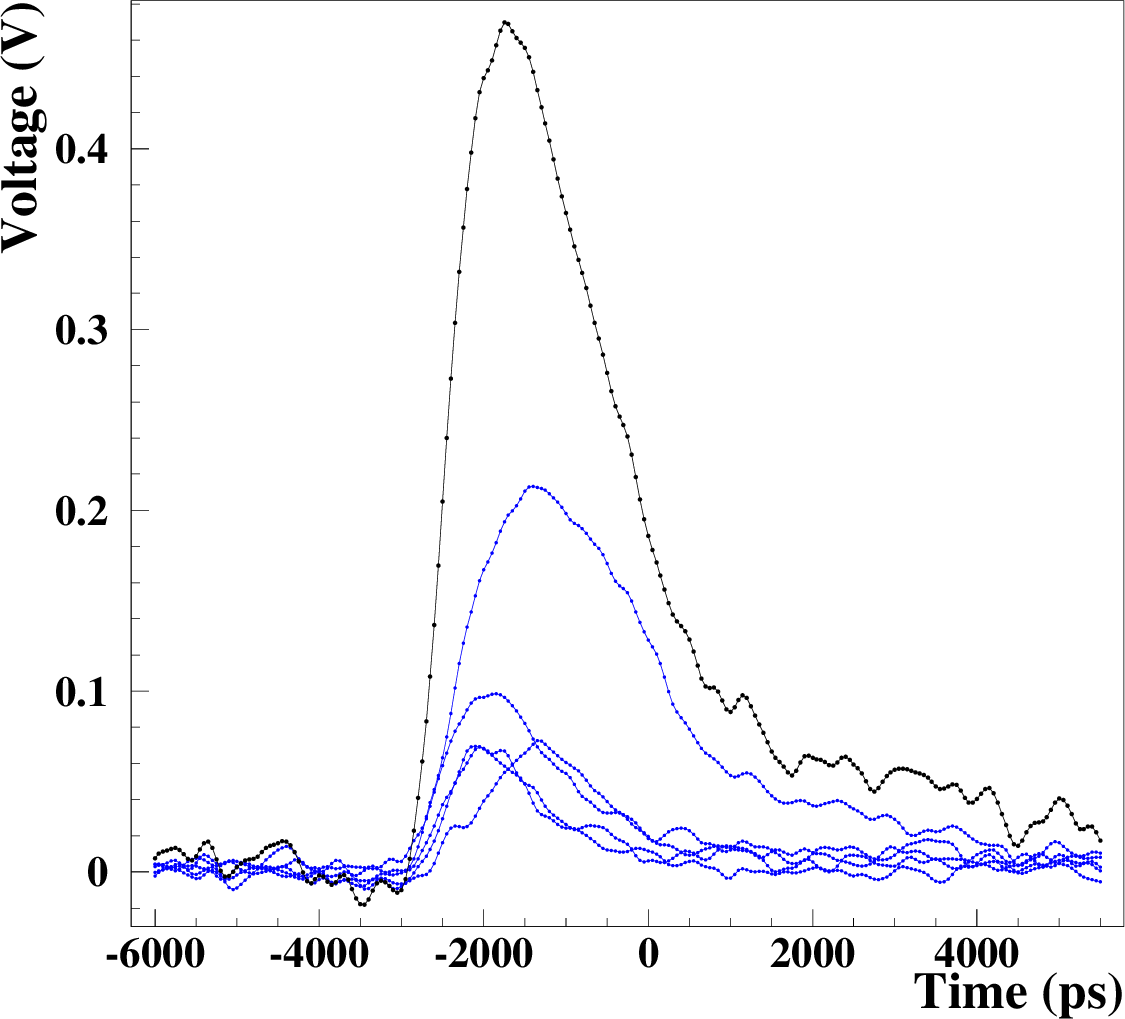}
		\caption{}
		\label{fig:spe-sum}
	\end{subfigure}
	\begin{subfigure}{0.5\textwidth}
		\centering
		\includegraphics[width=0.7\textwidth]{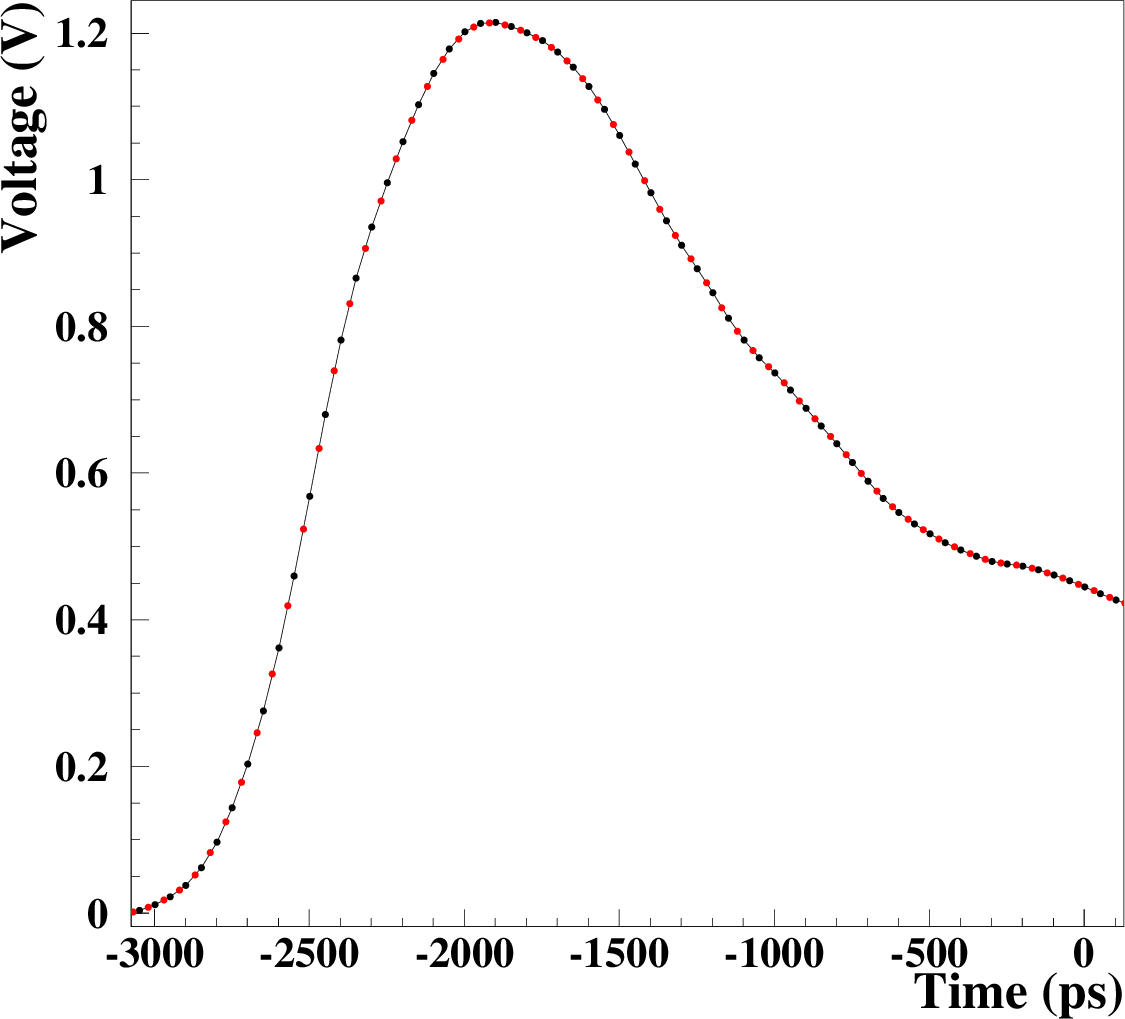}
		\caption{}
		\label{fig:wave_shift50}
	\end{subfigure}
	\caption{(a) Example of the simulated pulse of the PICOSEC-MM response to 5 photoelectrons (black), by summing randomly selected 5 waveforms of the SPE-set (blue). (b) Example of a simulated multi-pe pulse shifted within $\pm$50\,ps. Black points are the original digitizations of the waveform, while the red points are the shifted ones. The line connects the black digitization points.}
\end{figure}

Supposing that PICOSEC-MM (including the front-end electronics) is a linear device,  the detector response to N-photoelectrons should be emulated by the sum of N single-pe waveforms. For the simulation of waveforms like those comprising the EXP-set, we implement the following algorithm: 

A multiplicity, N of single-photoelectron (single-pe) signals is chosen according to a Poisson distribution with a mean value of 7.8. N waveforms are randomly selected among those of the SPE-set. \added{Each selected SPE waveform is time-aligned to a common reference (zero) using a 3$^{\mathrm{rd}}$-degree polynomial interpolation of the leading edge, and the aligned waveforms are summed to form an equivalent multi-photoelectron pulse. This exact alignment, however, artificially removes the natural timing spread that exists between pulses and the finite sampling-phase/trigger jitter of the digitizer: with perfect alignment, every summed pulse would have identical digit values at each sampling point. 
To restore realistic conditions, we introduce a stochastic time-shift: each summed pulse is shifted by a random Gaussian offset with standard deviation $\sigma=25\ \mathrm{ps}$ (corresponding roughly to $\pm 50\ \mathrm{ps}$ range), thereby emulating the combined effects of photoelectron arrival-time spread and digitizer/trigger sampling jitter.}  An example before and after applying this jitter is shown in Fig.~\ref{fig:wave_shift50}.

Using the bootstrap technique \cite{ramachandran2009mathematical} and the Poisson Distribution as a metric, we can generate random samples of single-pes to create simulated multi-pe pulses (approximately 100k events data set). However, this method has the disadvantage of introducing a correlation between the generated pulses since the final pulses share similar single-pe waveforms. 
\section{Validation of the Emulation Model}\label{sec:evaluation}
\begin{figure}[hbt!]
	\begin{subfigure}{0.5\textwidth}
		\centering
		\includegraphics[width=0.7\textwidth]{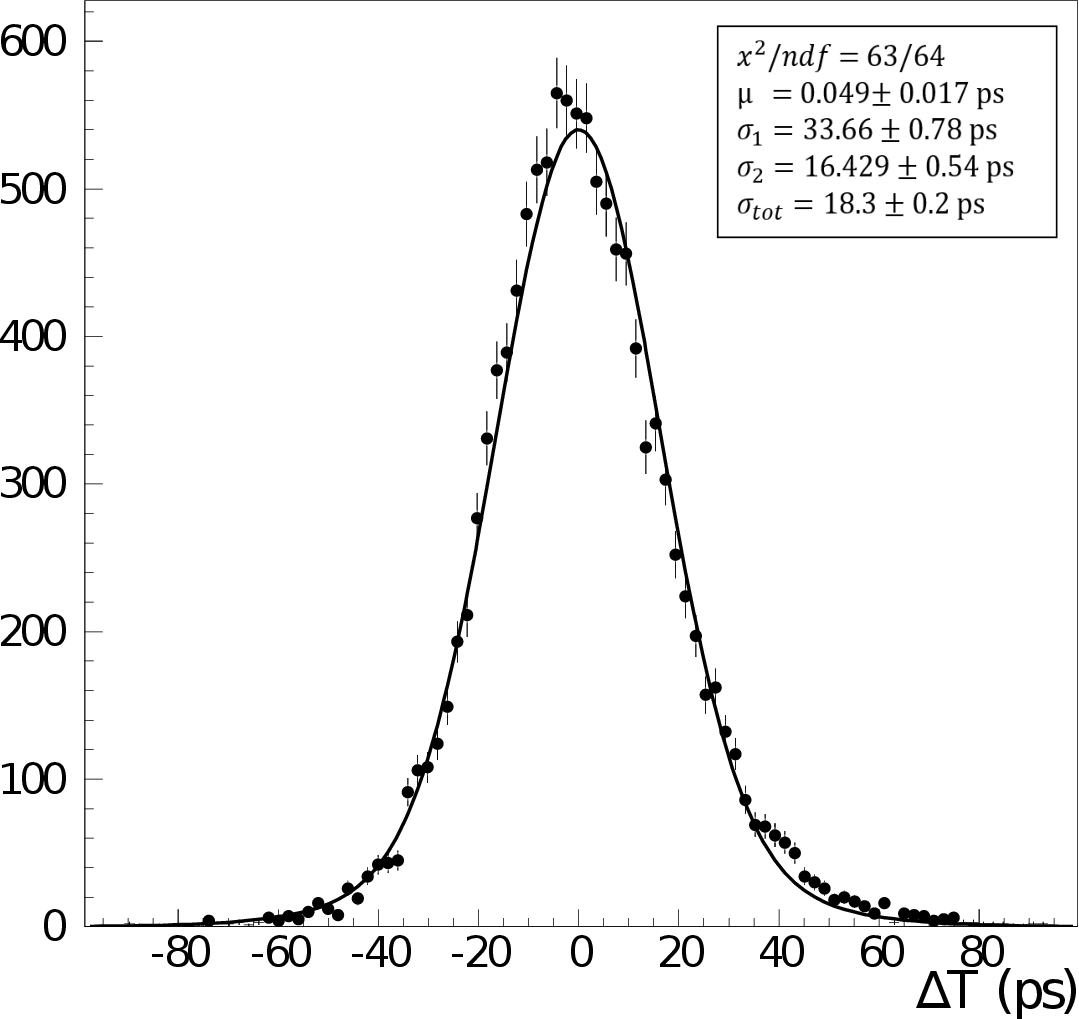}
		\caption{}
		\label{fig:EXP-resol}
	\end{subfigure}
	\begin{subfigure}{0.5\textwidth}
		\centering
		\includegraphics[width=0.7\textwidth]{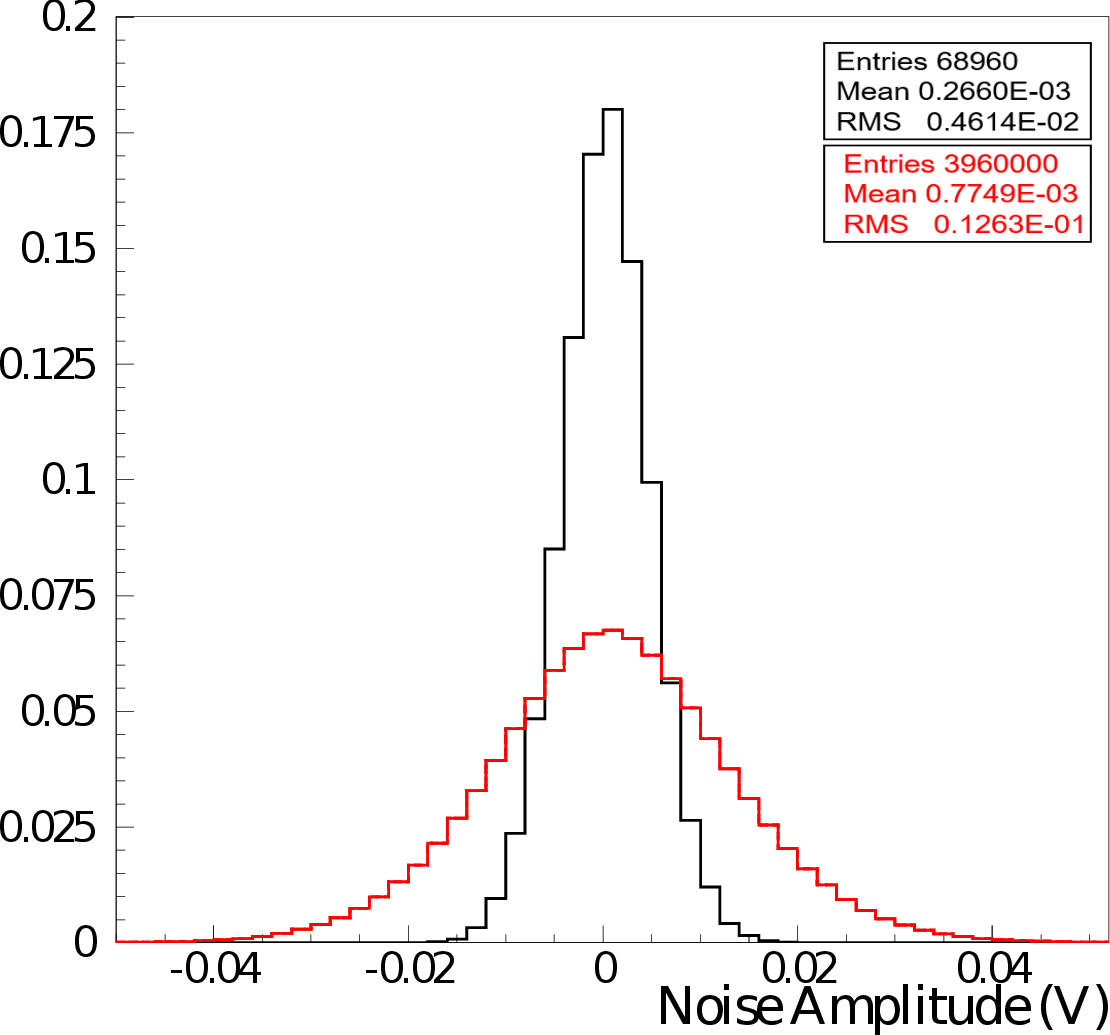}
		\caption{}	
		\label{fig:rms_noise_tot}
	\end{subfigure}
	\caption{(a) SAT distribution of the EXP-set using the CFD Technique, resulting to a global timing resolution of 18.3$\pm$0.2\,ps after time walk corrections and fitted with the weighted sum of two Gaussian distributions.  (b) Distribution of baseline RMS noise on single-pes(black) and emulated multi-photoelectrons(red).}
\end{figure}

The data set of emulated multi-pes waveforms is expected to be identical to the response of the PICOSEC-MM detector on the laser beam (EXP-set). 
%
and share the same timing properties as the EXP-set. EXP-set data have been fully analyzed offline, following the algorithm described in \cite{Aune:2020mwh}. After applying time walk corrections, parameterized as a function of the SAT to the e-peak charge, results in a timing resolution of 18.3$\pm$0.2\,ps, \added{shown in Fig.\ref{fig:EXP-resol}}.

%
\added{The full offline analysis of PICOSEC-MM waveforms requires the calculation and subtraction of the baseline offset, while the baseline RMS is also determined on an event-by-event basis~\cite{Bortfeldt2019}. For each waveform, the baseline is estimated within a signal-free time window of at least 80\,ns preceding the pulse. The amplitude distribution in this region is fitted with a Gaussian function, whose mean value defines the baseline offset and whose standard deviation corresponds to the baseline RMS. This procedure corrects baseline drifts and accounts for fluctuations in the electronic noise level on an event-by-event basis.}
For the SPE-set, the baseline RMS distribution is presented with black in Fig.\ref{fig:rms_noise_tot}. Thus, during the summing process, the generated pulses present an increasing RMS baseline noise (red-colored distribution in Fig.\ref{fig:rms_noise_tot}), analogous with the square root of the number of photoelectrons. 

The simulated pulses 
suffer from extra artificial noise, caused by the summing method of the single-pes pulses. For the simulation, data of the SPE-set were used with e-peak charge greater than 1\,pC, or amplitude of 30\,mV, to reduce the noise contribution. Unfortunately, this selection cut restricts the number of available events, hence the missing part of the spectrum needs to be replaced. This replacement is realized by scaling down larger pulses, in accordance with the Polya charge distribution.  

\subsection{Comparison of the Emulation Model with Experimental Data (EXP-set)}

\begin{figure}[hbt!]
	\begin{subfigure}{0.5\textwidth}
		\centering
		\includegraphics[width=0.7\textwidth]{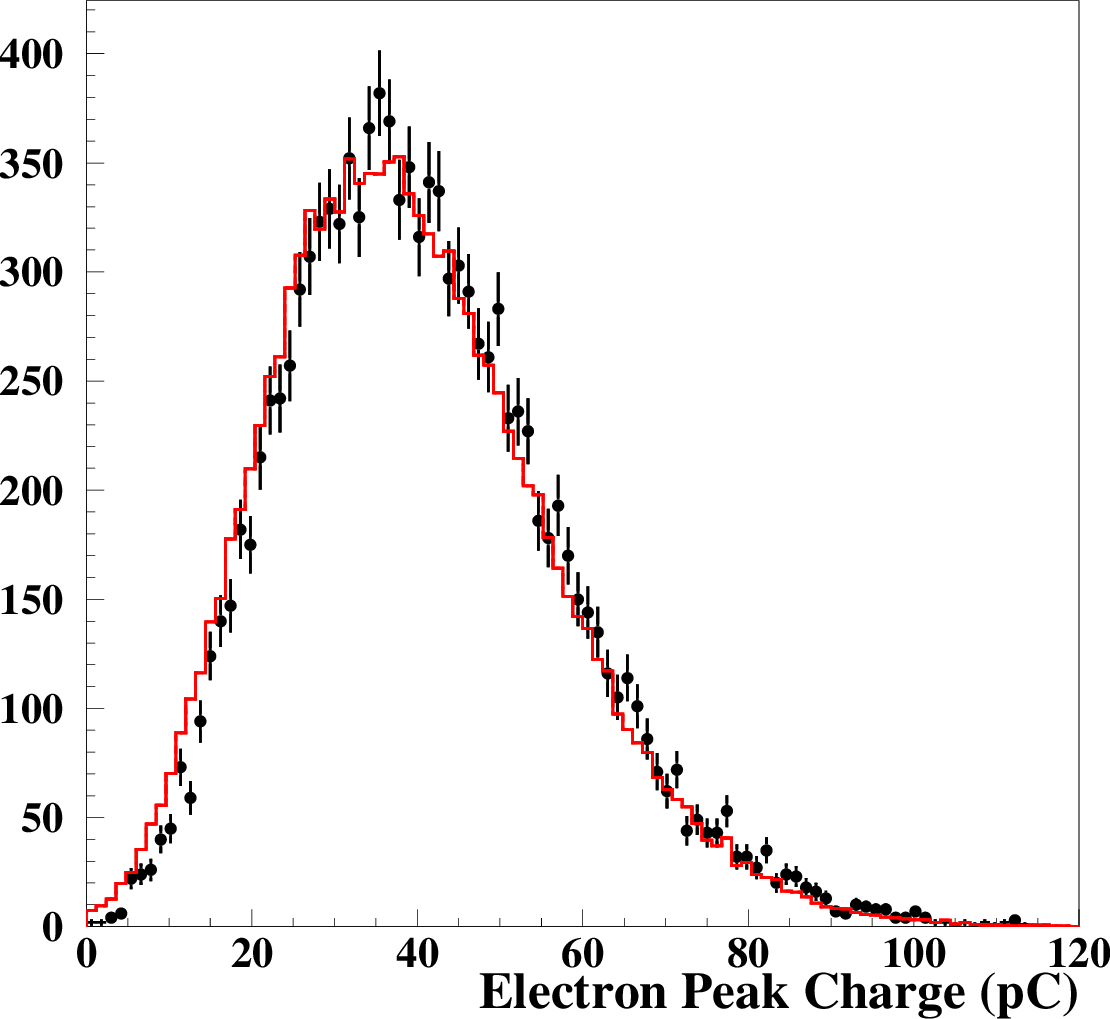}
		\caption{}
        \label{fig:charge_gpeak_simul}
	\end{subfigure}
	\begin{subfigure}{0.5\textwidth}
		\centering
		\includegraphics[width=0.7\textwidth]{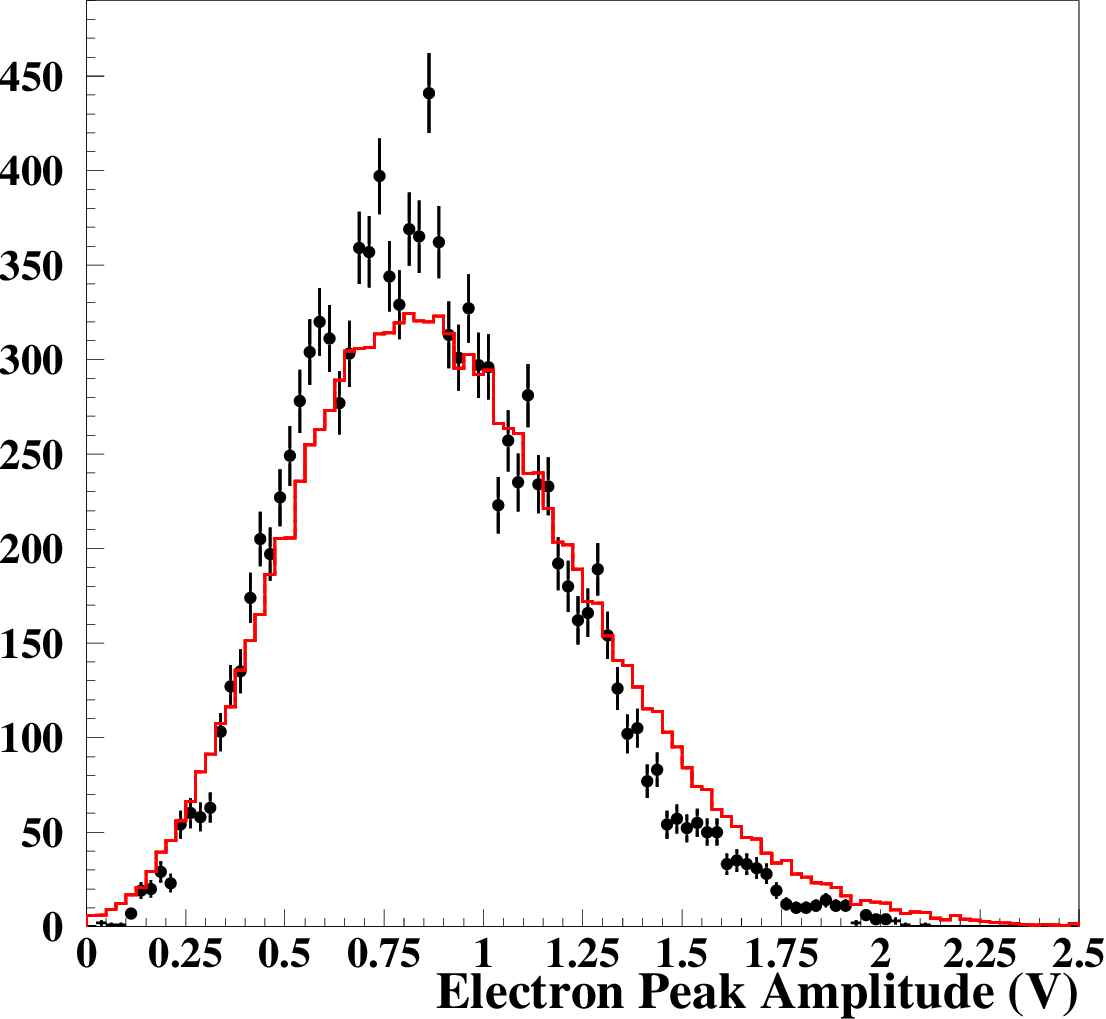}
		\caption{} 
        \label{fig:ampl_gpeak_simul}
	\end{subfigure}
	\caption{(a) E-peak Amplitude and (b) E-peak Charge Distributions for both the simulated-set (black) and the EXP-set (red) of waveforms.}
\end{figure}

As a cross-reference, to validate the simulated results, we compare them with the EXP-set. The distributions of e-peak charge and the peak amplitude are presented in Fig.\ref{fig:charge_gpeak_simul}, where black points are the simulated events and red are the experimental ones. 

Fig.\ref{fig:ampl_gpeak_simul} also provides evidence that EXP and simulated sets differ slightly regarding the e-peak amplitude distributions. Defining the maximum value for the amplitude of a signal requires the localization of the point with the highest value. It is obvious, that this measurement is biased from noise contribution. As shown (Fig.\ref{fig:rms_noise_tot}), simulated signals suffer from extra noise and as a result, the peak amplitude distribution is shifted to higher amplitude values. This constitutes a limitation of the model, which has to be considered in the full offline analysis of the simulation set. On the contrary, the e-peak charge is a more reliable parameter for the simulation, as it neutralizes noise effects and produces a similar distribution to the experimental data. 

\subsection{Evaluating the Timing Properties of the Emulation Model}\label{sec:timingsimul}

The evaluation of the timing properties of the model and their comparison with the EXP-set using the same offline analysis method, as performed in \cite{kallitsopoulou2021development}, will ensure the validity of the model. 

\begin{figure}[hbt!]
	\begin{subfigure}{0.5\textwidth}
		\centering 
		\includegraphics[width=0.7\textwidth]{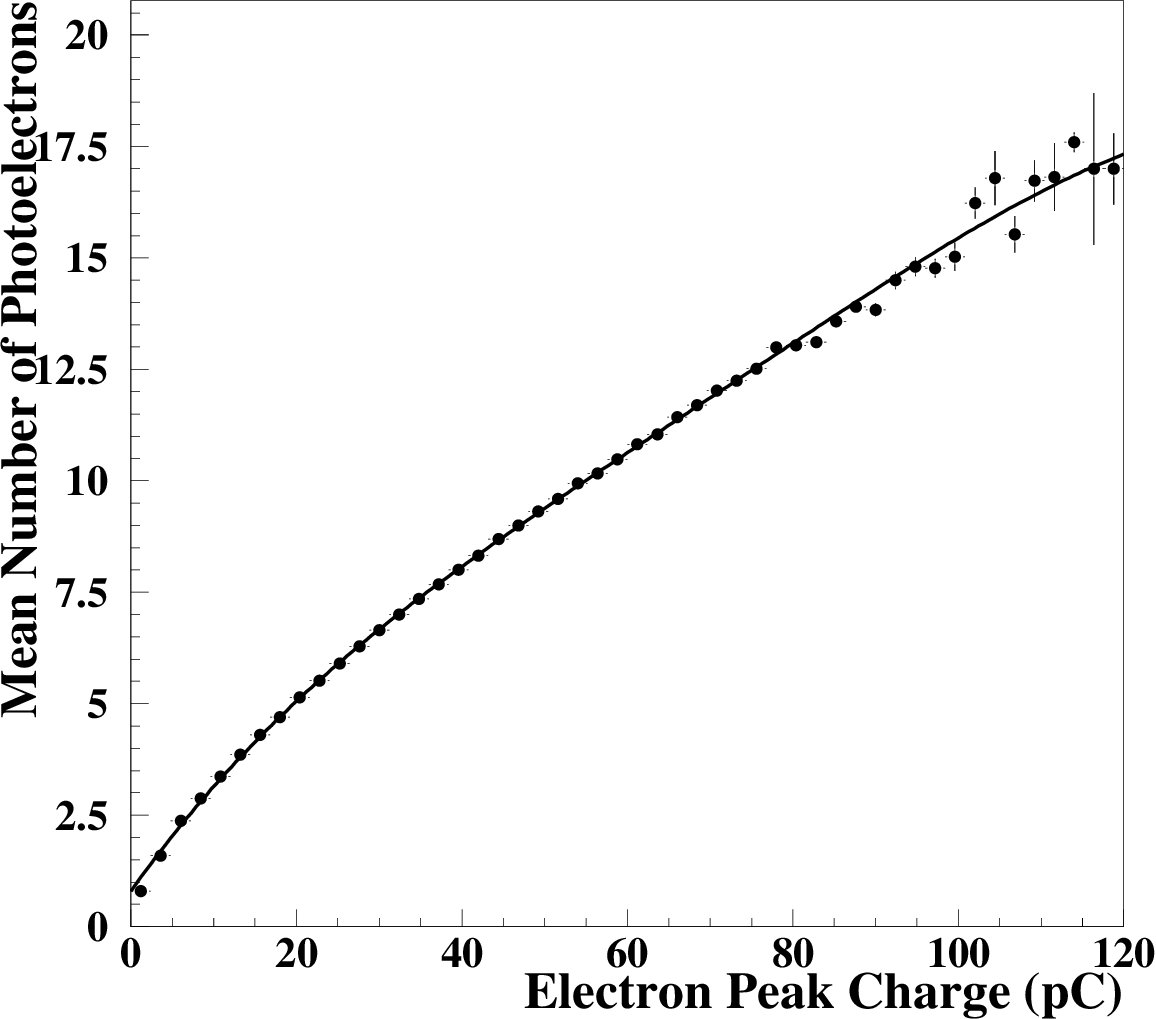}
		\caption{}\label{fig:meanpe_chagre}
	\end{subfigure}
	\begin{subfigure}{0.5\textwidth}
		\centering
		\includegraphics[width=0.7\textwidth]{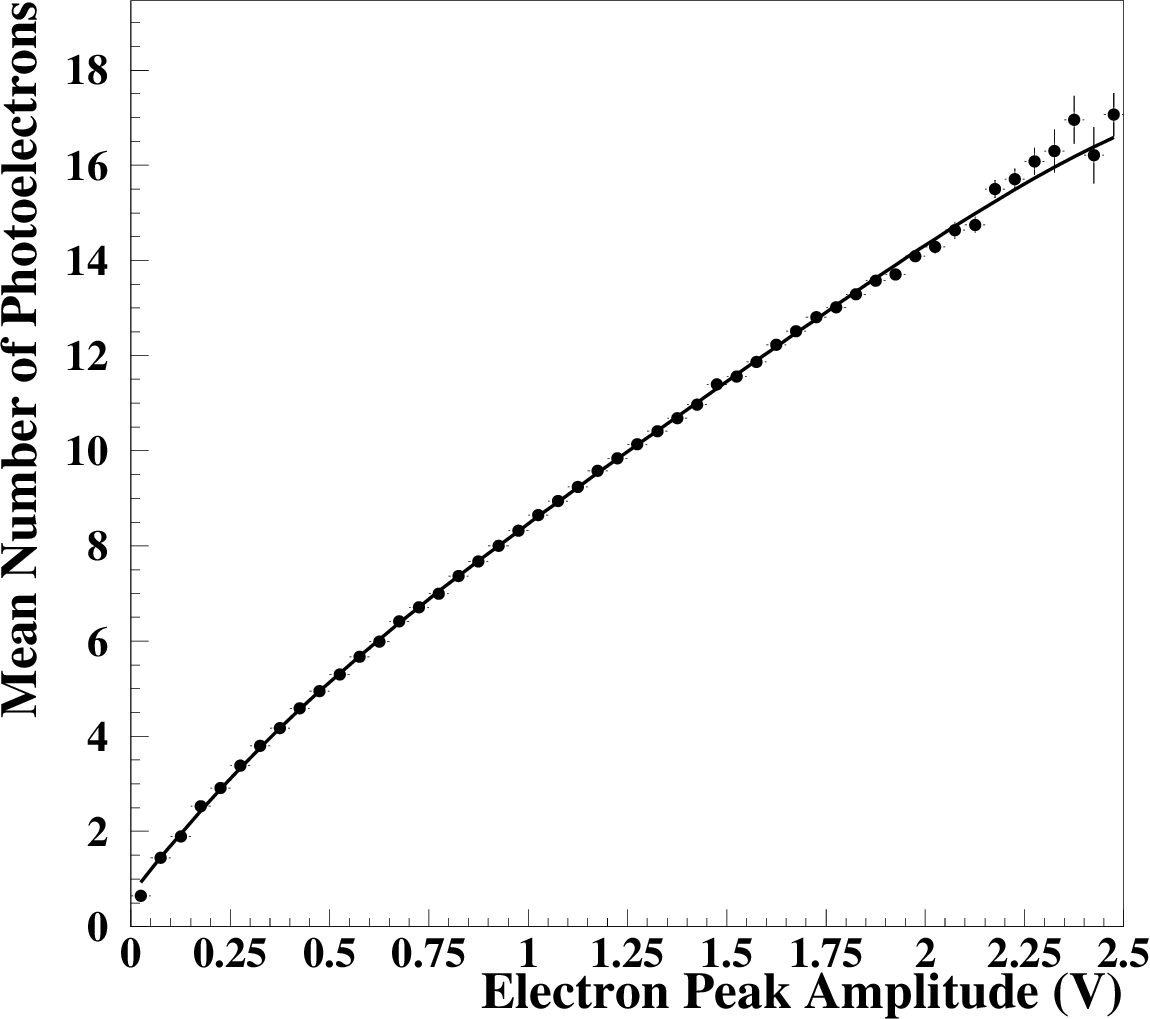}
		\caption{} 
		\label{fig:meanpe_gpeak}
	\end{subfigure}
	\caption{Correlation plot for the number of photoelectrons estimation relevant to e-peak size. (a) E-peak Charge distribution and (b) E-peak amplitude distribution. In both plots, the black solid line represents a $4^{rth}$ degree polynomial fit to the data points. }
	\label{fig:meanpe}
\end{figure}

\added{It is essential, for the analysis of the emulated pulses, to quantify the correlation between the e-peak size and the corresponding number of photoelectrons. To determine this relation, two distributions are constructed from the simulated dataset: (i) the total e-peak charge distribution, and (ii) the same distribution where each event is weighted by its number of photoelectrons, $N_{\mathrm{pe}}$. The ratio of the weighted to the unweighted distributions provides the mean number of photoelectrons as a function of the e-peak charge, as shown in Fig.~\ref{fig:meanpe_chagre}. The statistical uncertainties are derived from the weighted charge distribution and scaled with $N_{\mathrm{pe}}^{2}$. A similar analysis can be performed using the e-peak amplitude instead of the charge (Fig.~\ref{fig:meanpe_gpeak}). Each resulting dependence is parameterized with a fourth-degree polynomial for subsequent use in the waveform reconstruction.}



\begin{figure}[hbt!]
	\begin{subfigure}{0.5\textwidth}
		\centering 
		\includegraphics[width=0.7\textwidth]{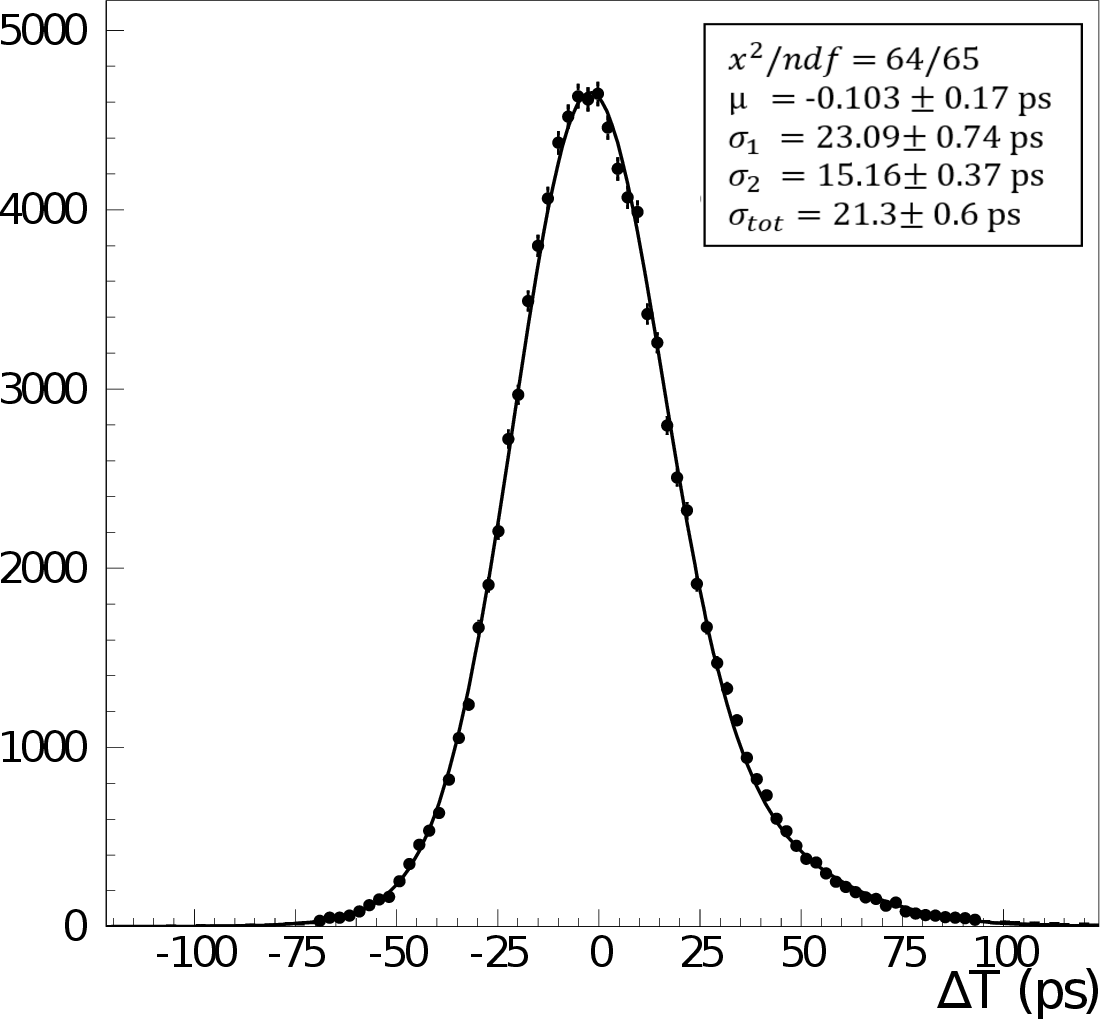}
		\caption{}\label{fig:simul_dt_corr}
	\end{subfigure}
	\begin{subfigure}{0.5\textwidth}
		\includegraphics[width=0.7\textwidth]{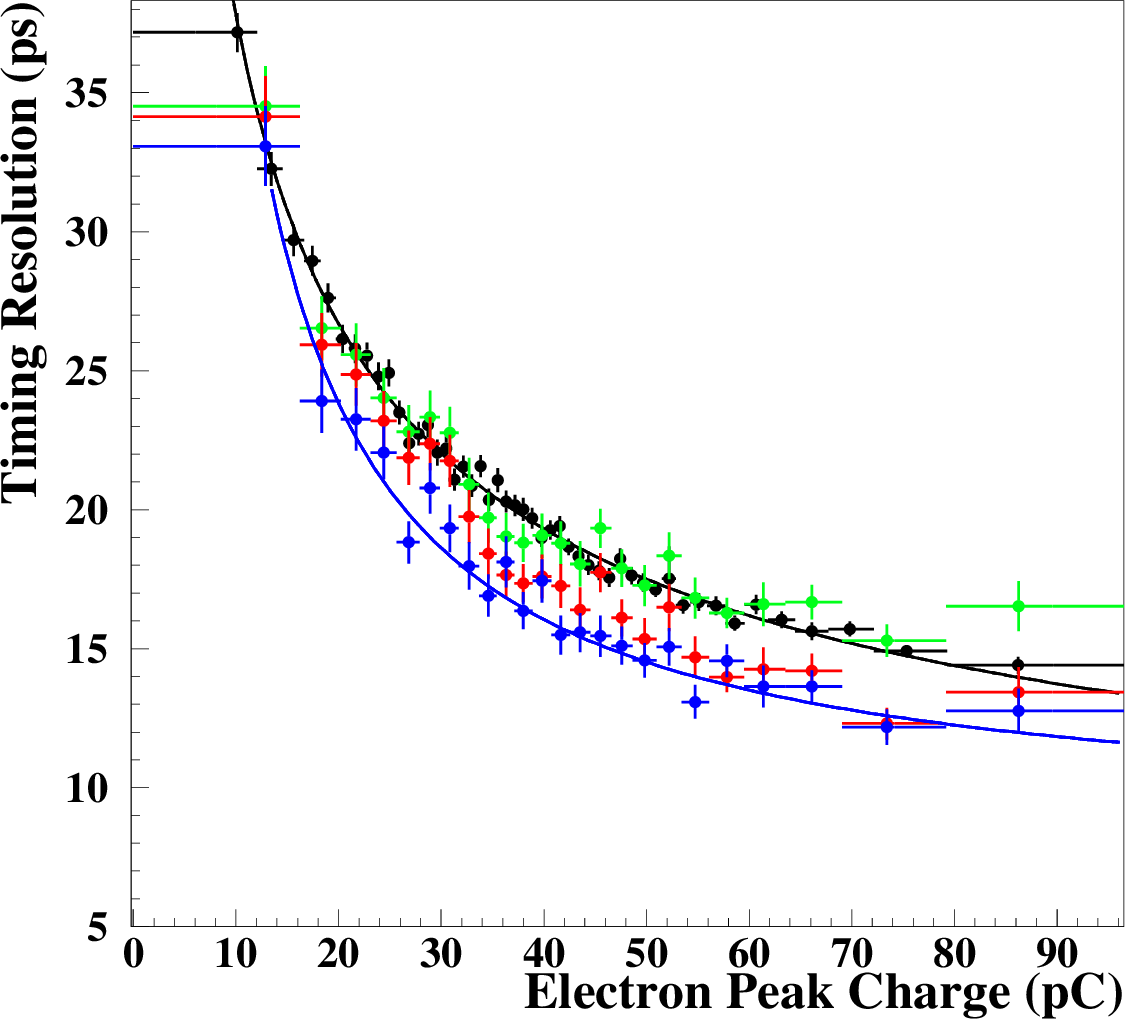}
		\caption{} 
		\label{fig:simul_resol_tot}
	\end{subfigure}
	\caption{(a) SAT distribution of the model, after time walk corrections, resulting to a timing resolution of 21.3$\pm$0.6\,ps, fitted with the weighted sum of two Gaussian distributions. (b) Resolution as a function of e-peak charge of: Simulated data (black), EXP-set (blue), EXP-set shifted due to extra RMS baseline noise according to the number of photoelectrons (red) and EXP-set additionally shifted due to photodiode's precision as a function of the number of photoelectrons (green). Lines correspond to fit with power law.}
\end{figure}

The full offline timing analysis of the simulated pulses, using the calibration curves for time walk correction, results in a timing resolution of 21.3$\pm$0.6\,ps, as it can be seen in Fig.\ref{fig:simul_dt_corr}. The timing resolution of the model is presented with black points in Fig.\ref{fig:simul_resol_tot}. The timing precision of the model is lagging by 3\,ps compared to EXP-data, shown as blue points in Fig.\ref{fig:simul_resol_tot}. The deteriorated resolution of the simulated data is attributed to two main sources: the baseline RMS noise and the synchronization limitations from using a common time reference. The sum of $\bar{N} = 7.8$ single-photoelectron pulses accumulates additional baseline RMS noise, while the timing accuracy of the reference device restricts resolution based on the number of photoelectrons, \textit{N}. Though noise accumulation constrains the method for developing new pulses, these two error components can be added to the EXP-data, making both sets subject to the same noise. 

To account for additional noise in the EXP-set, each data point is shifted event by event based on the number of photoelectrons, which provide signal information. The extra RMS noise, represented as $\sigma_1 = \sqrt{N_{pe}}\cdot \sigma_{1pe}$, where $\sigma_{1pe} = 0.0046$ (estimated in the Fig.\ref{fig:rms_noise_tot}), is added by shifting waveforms through a Gaussian with mean equal to zero and $\sigma = \sigma_1$. This adjustment shifts the EXP-data points toward the simulated ones, as seen with the red points in Fig.\ref{fig:simul_resol_tot}. Additionally, the photodiode's precision contributes further error based on the number of photoelectrons, $N_{pe}\cdot\sigma_{phd}$, which is combined quadratically with the RMS noise, ultimately shifting the red points to become the green ones, as depicted in the same figure. The total effect results in a variance of : 

\begin{equation}
	V[Data] = V[Data_{noise}] + V[phd] = N_{pe}\cdot \sigma_{1pe}^2 + N_{pe}^2\cdot\sigma_{phd}^2 
\end{equation} 
where the error of photodiode's precision ($\sigma_{phd}$) is approximately 3\,ps\cite{sohl:tel-03167728}. 

\section{Using the Emulation Model to Train an Artificial Neural Network}\label{sec:training}

Estimating timing resolution using Artificial Neural Networks can be highly beneficial, enabling real-time or near-real-time timing analysis. Additionally, it can easily be combined with other data analysis techniques, such as machine learning algorithms for signal classification or data filtering. It can become an important tool to enhance overall timing resolution estimation. 

\subsection{Validation of the Artificial Neural Network's Performance}

\begin{figure}[hbt!]
	\begin{subfigure}{0.5\textwidth}
		\centering 
		\includegraphics[width=0.7\textwidth]{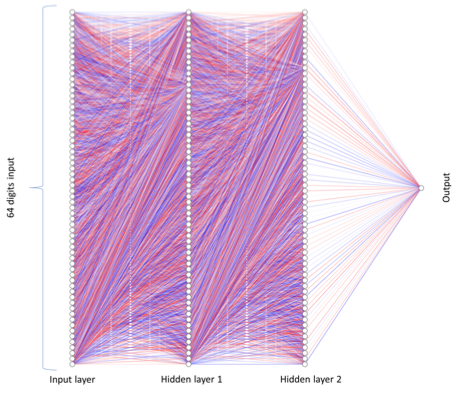}
		\caption{}\label{fig:neural_sketch}
	\end{subfigure}
	\begin{subfigure}{0.5\textwidth}
		\centering 
		\includegraphics[width=0.7\textwidth]{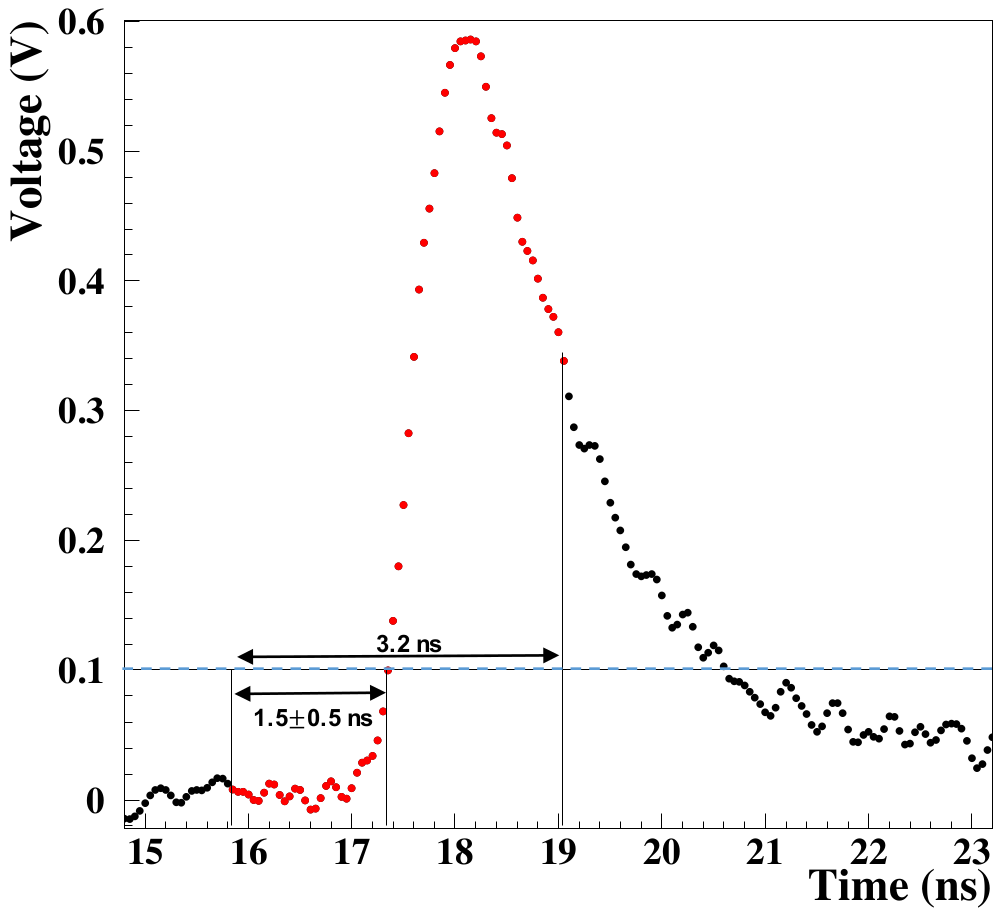}
		\caption{}\label{fig:nn_wave}
	\end{subfigure}
	\caption{(a) Artificial Neural Network architecture comprising an input layer, two hidden layers, and one output layer. The number of input and hidden nodes is defined by the number of selected digitizations of the e-peak waveform. (b) Typical PICOSEC-MM digitized waveform. Red points denote the digitized information presented to the ANN, in this case -the 64 digits around the trigger point- correspond to 3.2\,ns time interval.}
\end{figure}

In the present study, the Neural Network implementation was based on a Feed-Forward NN architecture \cite{goodfellow2016deep}.  In this context, the ANN consists of an input layer, two hidden layers with 64 neurons each, and an output layer. This model has been found to give adequate results and can be visualized in Fig.\ref{fig:neural_sketch}. For all nodes, \emph{ReLU} activation function \cite{agarap2019deep} was used apart from the output layer, an essential part of regression tasks. In parallel, the mean squared error was used as a loss function.  

In the present study, two training scenarios were performed: 
\begin{itemize}
	\item The predictive capabilities of the ANN were first studied with experimental data using the k-fold cross-validation technique, which was already mentioned in the introduction of Section \ref{sec:simulation} \cite{francois}.
	\item After the model implementation, a large data-set of pulses is available. The training procedure using the model is analogous to the previous case, without the folding. The model was later on tested using the total data set of the experimental set. 
\end{itemize} 

The training set used in each case was split into training and validation sets, with a ratio of 0.2. The model was trained for 10000 epochs. At each iteration, the model is evaluated on an unseen validation set, and its weights are updated only if the cost function shows improvement on this validation set.

Waveforms fed in the ANN were like the one seen in Fig.\ref{fig:nn_wave}. \added{Red points denote the digitized information presented to the ANN, in this case the 64 digits around the trigger point-corresponding to 3.2\,ns time interval.} A 100\,mV threshold trigger is chosen, giving a selection criterion to the waveforms and providing a timestamp. This timestamp is used to determine the starting time point which is 1.5\,ns before the crossing point with the threshold. A shift of the starting time with a Gaussian with $\sigma$=500\,ps, for the ANN to be unbiased of the selected starting point is applied. Subsequently, this will define the total time window of 3.2\,ns (64-bit digitization with a sampling rate of 50\,ps) given to the input layer of the ANN (red-colored points). Various input lengths were tested, and it was found that the optimum performance was achieved when the input layer contained mostly the leading edge. Moreover, during the training period, the reference time was provided to the network as the target value. 					 

\begin{figure}[hbt!]
	\begin{subfigure}{0.5\textwidth}
		\centering 
		\includegraphics[width=0.7\textwidth]{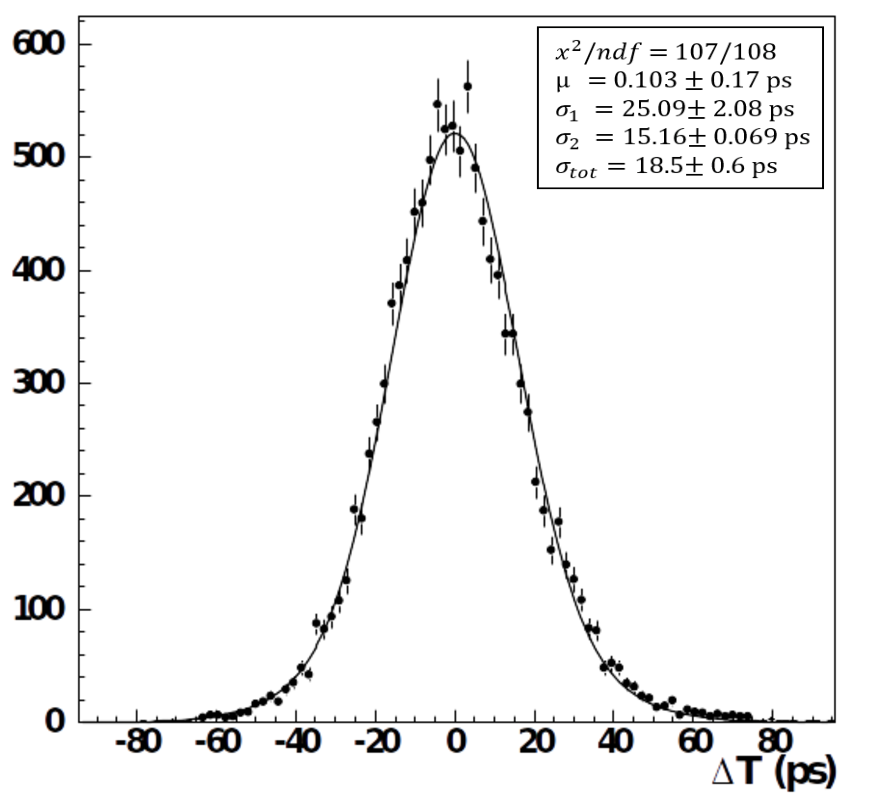}
		\caption{}\label{fig:nn_cdf}
	\end{subfigure}
	\begin{subfigure}{0.5\textwidth}
		\centering 
		\includegraphics[width=0.75\textwidth]{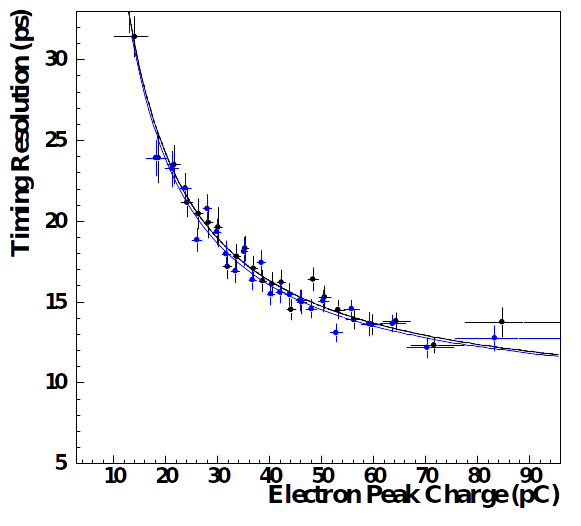}
		\caption{}\label{fig:nn_resol_qe}
	\end{subfigure}
	\caption{(a)The PICOSEC-MM Signal Arrival Time, evaluated by the Neural Network, fitted with a Gaussian distribution with RMS that determines the global timing resolution at 18.5 $\pm$ 0.6 \,ps. (b) Timing Resolution as a function of e-peak charge. The ANN results (blue points) are compared with those of the CFD Technique (black points).}\label{fig:gauss_nn}
\end{figure}

Using the method of training with k-fold cross-validation, only with experimental data (EXP-set, with 7.8\,pes), the ANN results in a timing resolution of 18.5$\pm$0.6\,ps, shown in Fig.\ref{fig:nn_cdf}. In Fig.\ref{fig:nn_resol_qe}, the timing resolution using the ANN is compared with the results of the full offline analysis using the CFD technique. The similarity confirms the validity of the ANN timing method performance.

\subsection{Validation of the ANN’s Unbiased Performance}

\begin{figure}[hbt!]
	\centering 
	\includegraphics[width=0.7\textwidth]{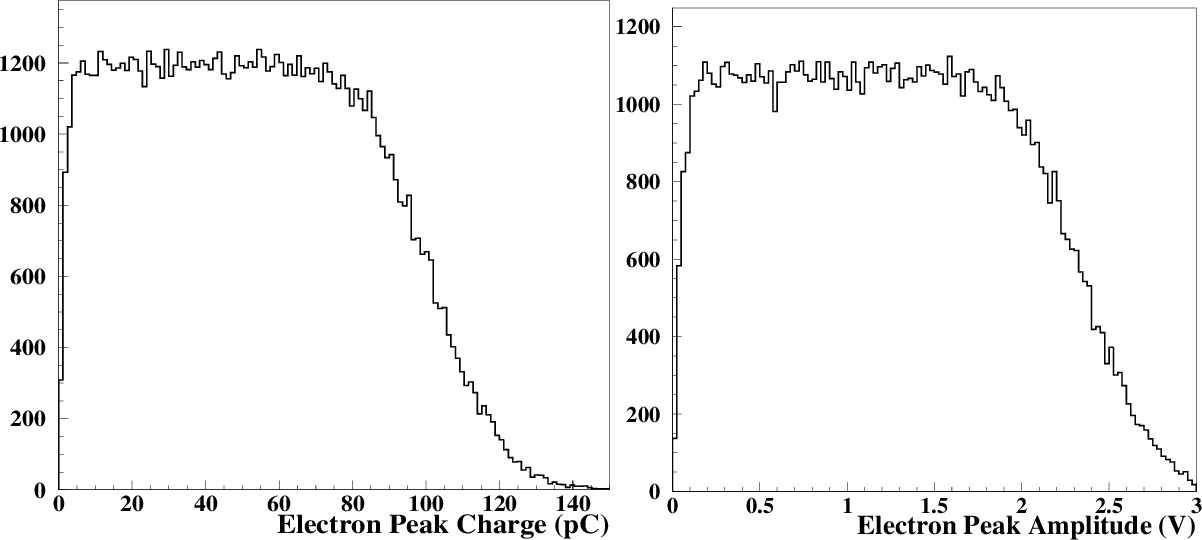}
	\caption{{Training of the ANN with uniform distribution of e-peak amplitude (right), and e-peak charge (left).}\label{fig:flat_nn}}
\end{figure}

The second training procedure approach uses as a training set the simulated pulses. Aiming for a training session unbiased of the e-peak size, the model was developed with uniform charge and amplitude distributions\footnote{The uniform region of distribution represents the vast majority of the EXP-set signals}, as shown in Fig.\ref{fig:flat_nn}. Again, the ANN reaches the same timing accuracy as the full signal processing analysis of 18.5 $\pm$ 0.6 \,ps.

\begin{figure}[hbt!]
	\begin{subfigure}{0.5\textwidth}
		\centering 
		\includegraphics[width=0.7\textwidth]{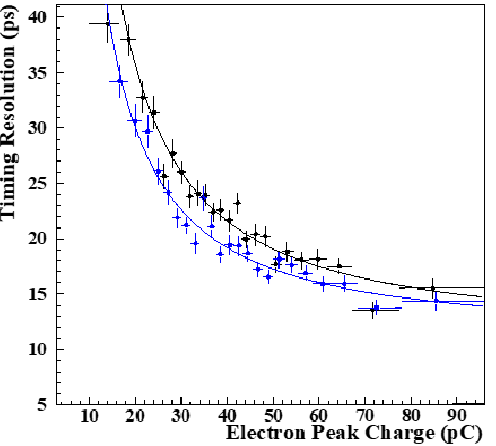}
		\caption{}\label{fig:NN_tests_a}
	\end{subfigure}
	\begin{subfigure}{0.5\textwidth}
		\centering 
		\includegraphics[width=0.75\textwidth]{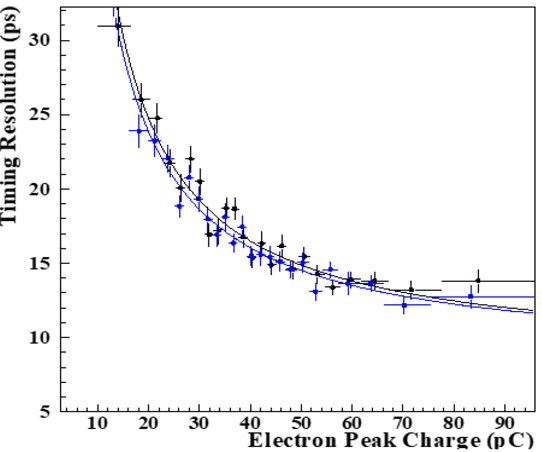}
		\caption{}\label{fig:NN_tests_b}
	\end{subfigure}
    \begin{subfigure}{0.5\textwidth}
		\centering 
		\includegraphics[width=0.75\textwidth]{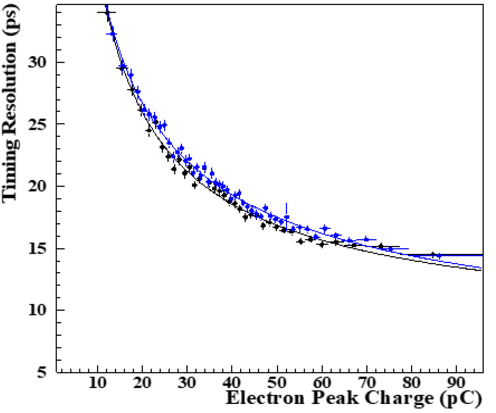}
		\caption{}\label{fig:NN_tests_c}
	\end{subfigure}
    \begin{subfigure}{0.5\textwidth}
		\centering 
		\includegraphics[width=0.75\textwidth]{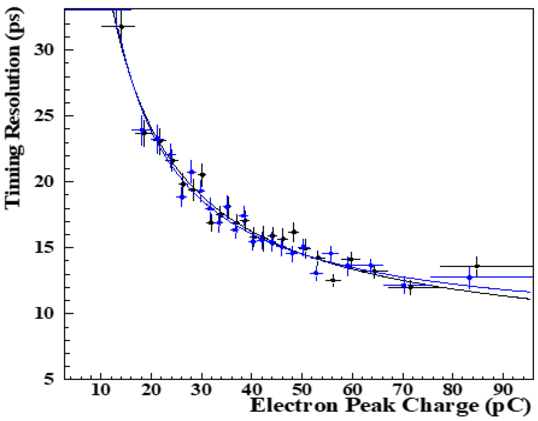}
		\caption{}\label{fig:NN_tests_d}
	\end{subfigure}
	\caption{Timing resolution as a function of e-peak Charge, for the four different tests passed from the ANN(black points), in comparison with the full offline analysis using CFD Technique(blue points). (a) Adding noise of 30\,mV at EXP-data-set. (b) Test of the  ANN with 1.25ns timestamp. (c) Evaluation of the performance of ANN in simulated data both for train and test processes. (d) Test of the ANN with fewer digitization points (16 instead of 64) as input layer.}\label{fig:nn_tests}
\end{figure}

Since there is strong evidence that ANN works properly, a variety of tests performed to prove its unbiased and consistent operation. It has been shown that the injection of noise to input training data can act as a regularization technique, preventing the ANN from overfitting. For this reason, a 15\,mV Gaussian noise was added to the training phase, resulting in better performance of the neural network algorithm and leaving unaffected the timing resolution. Additionally, aiming to understand whether or not the ANN performs with respect to the signal processing algorithm, a significant amount of noise was added to the testing data-set (EXP-set). Consequently, on an event-by-event basis, the EXP-set was shifted within a Gaussian with mean zero and RMS 30\,mV. The full offline analysis of the shifted EXP-set, results in a resolution of 23.6$\pm$0.5\,ps. For the same data, used as a testing set, the ANN results to 24.7$\pm$0.6\,ps (Fig.\ref{fig:NN_tests_a}). 

The next evaluation test uses only the simulated data produced by the model. The simulated data set produced with uniform charge distribution was used as a training set, then the simulated data set was used as a testing set according to the Polya charge distribution of the EXP-set. In this way, the ANN results in a timing resolution of 21.3$\pm$0.6\,ps, similar to the one achieved with the full offline analysis (Fig.\ref{fig:NN_tests_c}). 

Another test is to provide fewer input points of each waveform on the ANN training session, i.e. 1.25\,ns instead of 1.5\,ns (with the same 50\,ps spacing). Additionally, the jitter of the timestamp is changed to 100\,ps. This procedure results in a timing resolution of 18.4$\pm$0.6\,ps, similar to the CFD analysis result (Fig.\ref{fig:NN_tests_b}). The last test utilizes of different sampling rate with a digitization of 200\,ps instead of 50\,ps, resulting in 16 digits as the input layer of the ANN. This demands also a different training set, with the same digitization before the ANN train (i.e the multi-pes waveforms will have a wider digitization interval). Performing the full offline analysis on this sample results in a timing precision of 19.01$\pm$0.5\,ps. Respectively, the ANN achieves 19.02$\pm$ 0.5\,ps (Fig.\ref{fig:NN_tests_d}). 

    

\section{Conclusions}\label{sec:conclusion}

This study demonstrates the development of advanced signal processing techniques for online, precise timing. The EXP-data set was collected in Laser Beam Tests and consisted of PICOSEC-MM signals with an average of 7.8 photoelectrons (pes). The standard CFD timing technique was used to analyze these signals, resulting in a resolution of 18.3$\pm$0.2\,ps, which served as a benchmark for evaluating the results of newly developed timing algorithms.

One such alternative algorithm is the Constant Threshold Discrimination timing technique, which uses the charge over threshold as the variable for time walk correction. This technique provides results in excellent agreement with the reference method and can easily be integrated into electronic devices, such as the NINO chip accompanied by ADC measurements.

Another technique proposed in this study is an Artificial Neural Network (ANN) for real-time signal processing, which holds great promise for fast event selection. To train the ANN, a model was created due to the need for a large training sample. After verifying that the model was consistent with the detector's response to many photoelectrons, it was used to train the ANN to extract timing information. The ANN passed all tests with high precision timing performance, proving that it is a reliable and unbiased tool, despite the model used as the training set consisting of waveforms that are not completely uncorrelated. The emulated data proved sufficient for the ANN to learn the signal analysis procedure for precise timing. Significantly, this first implementation was developed in a regime where the detector design ensured minimal time-walk effects---only a few picoseconds in the thin-gap prototype \cite{utrobicic2024singlechannelpicosecmicromegas}.\added{ Under such conditions, the ANN approach is highly effective because the uniformity of the waveform shapes allows the network to focus on subtle temporal features rather than compensating for amplitude-dependent variations. In contrast, for detectors or operating conditions where time-walk effects are more pronounced, additional corrections or training strategies may be required, as the same method might no longer provide an accurate approximation, potentially limiting its timing resolution.}


It should also be emphasized that the ANN technique presented here represents only the first architecture tested within this framework. Other ANN-based approaches have already been developed \cite{Tsiamis:359504, charzianagnostou:343732} and will be reported in forthcoming publications. These complementary efforts highlight the ongoing progress in applying machine learning to PICOSEC timing, suggesting that further improvements in architecture and training strategies may unlock even higher performance.

Overall, this study highlights how both analytical signal processing algorithms and ANN-based methods can achieve precise timing resolution with PICOSEC-MM detectors. While analytical techniques provide robust and hardware-friendly solutions, ANN approaches hold great promise for flexible, real-time applications---particularly when time-walk effects are minimized.


\acknowledgments

This research work was co-funded by the Hellenic Foundation for Research and Innovation (HFRI) under the HFRI PhD Fellowship grant (Fellowship Number: 6741). Part of the work was also supported by the Deutsche Forschungs-gemeinschaft (DFG, German Research Foundation) under Germany’s Excellence Strategy – EXC 2121 ‘‘Quantum Universe’’ – 390833306. 
The authors would like to thank Mr. T. Gustavson and the IRAMIS facility of CEA Saclay, as well as the CERN-RD51 PICOSEC Micromegas collaboration, for providing experimental data from the PICOSEC detector.



\bibliographystyle{JHEP}
\bibliography{main.bib}

\end{document}